\documentclass[%
 aip,twocolum,
 amsmath,amssymb,
reprint,%
]{revtex4-2}

\usepackage{subfigure}
\usepackage{graphicx}
\usepackage{dcolumn}
\usepackage{bm}

\usepackage[utf8]{inputenc}
\usepackage[T1]{fontenc}
\usepackage{mathptmx}
\usepackage{etoolbox}
\usepackage{lipsum}

\usepackage{url}
\usepackage{verbatim}
\usepackage[pdftex,bookmarksnumbered]{hyperref} 

\makeatletter
\def\@email#1#2{%
 \endgroup
 \patchcmd{\titleblock@produce}
  {\frontmatter@RRAPformat}
  {\frontmatter@RRAPformat{\produce@RRAP{*#1\href{mailto:#2}{#2}}}\frontmatter@RRAPformat}
  {}{}
}%
\makeatother

\begin{document}

\title[Equilibrium theory of bidensity particle-laden suspensions in thin-film flow down a spiral separator]{Equilibrium theory of bidensity particle-laden suspensions in thin-film flow down a spiral separator}

\author{Lingyun Ding}
\email{dingly@ucla.edu}
\affiliation{ Department of Mathematics, University of California, Los Angeles, California 90095, USA}

\author{Sarah C. Burnett}
\email{burnett@math.ucla.edu}
\affiliation{ Department of Mathematics, University of California, Los Angeles, California 90095, USA}

\author{Andrea L. Bertozzi}
\email{bertozzi@math.ucla.edu}
\affiliation{ Department of Mathematics, University of California, Los Angeles, California 90095, USA}

\date{\today}
\begin{abstract}
Spiral gravity separators are designed to separate multi-species slurry components based on differences in density and size. Previous studies\cite{lee2014behavior, arnold2019particle} have investigated steady-state solutions for mixtures of liquids and a single particle species in thin-film flows. However, these models are constrained to single-species systems and cannot describe the dynamics of multi-species separation. In contrast, our analysis extends to mixtures containing two particle species of differing densities, revealing that they undergo radial separation—an essential mechanism for practical applications in separating particles of varying densities. This work models gravity-driven bidensity slurries in a spiral trough by incorporating particle interactions, using empirically derived formulas for particle fluxes from previous bidensity studies on inclined planes.\cite{wong2016conservation} Specifically, we study a thin-film bidensity slurry flowing down a rectangular channel helically wound around a vertical axis. Through a thin-film approximation, we derive equilibrium profiles for the concentration of each particle species and the fluid depth. Additionally, we analyze the influence of key design parameters, such as spiral radius and channel width, on particle concentration profiles. Our findings provide valuable insights into optimizing spiral separator designs for enhanced applicability and adaptability.

\end{abstract}

\maketitle

\section{Introduction \label{sec:Introduction}}

A spiral separator is a contemporary device known for its high capacity and cost-effectiveness in wet separation of particles based on their specific gravity. Also known as a spiral concentrator and gravity separator, it comprises a helical trough wrapped around an upright column. The slurry (or pulp), containing a mixture of particles and liquid, is introduced at the top of the spiral at a predetermined rate. As the slurry descends, the particles in the suspension aggregate toward the inner wall, closer to the center axis.
Centripetal forces, particle-particle interactions, and hydrodynamics cause the particles to concentrate by their densities at different radial positions within the channel, which results in a banded separation of the different particles at the outlet of the spiral separator. Adjustable splitters within the apparatus allow for further segregation of the sorted material into dedicated streams. In this paper, we conduct a theoretical analysis of slurries with two particle densities in the equilibrium state, aiming to predict their final distribution profile at the exit, where separation into distinct streams will occur.

Wet spiral concentrators are common in the coal processing and other mining industry, particularly for separating concentrations of heavy minerals such as gold, tin, tungsten, among others.\cite{Kernpatent, WrightpatentEU, giffardpatent, WrightpatentUS, weldon1997fine, holland1991particle} Wet separation is preferable to dry separation for hazardous material such as asbestos. Mixing the minerals with liquid for the separation process can keep harmful particles from becoming airborne.\cite{WrightpatentEU} The wet separation processing occurs in the food industry. For example, denser starch particles are separated from gluten to be removed from flour by using centrifuges and hydroclones.\cite{sayaslan2004wet} Centrifugal devices in bioengineering also implement density-based chromatography to separate components in mixtures.\cite{majekodunmi2015review}

The dynamics of particle-laden substances in a spiral separator are complex. For instance, as observed in the centrifuge device, heavier particles experience a stronger centrifugal force, causing them to migrate away from the center axis. Displaced smaller particles will migrate towards the inner axis.\cite{majekodunmi2015review} Therefore, one might expect a similar outcome in a spiral separator, with heavier particles concentrating at the outer sidewall. There is an application of spiral separators demonstrating this type of separation, although it is uncommon. During the transient descent of sand and similar particles, heavier particles move to the outer side of the channel, while lighter particles move to the inner side and escape through ports to be sorted.\cite{Kernpatent} Surprisingly, real-world spiral separators often defy intuitive expectations.\cite{boisvert2023axial, WrightpatentEU, giffardpatent, WrightpatentUS, weldon1997fine} For example, in coal benefication, coal is separated from impurities like silica and carbonaceous shale. In a spiral seperator, slurries exhibit a behavior where denser materials (coal) tend to travel along the inner region, next to the center axis, while lighter particles (silica) migrate on a path that is further from the inner axis.\cite{weldon1997fine} This seemingly counterintuitive behavior finds support in the current theoretical framework, which predicts secondary flows within the spiral channel that cause particles to accumulate on the inner side.\cite{lee2014behavior,arnold2019particle}

In light of these findings, a nuanced understanding of the behavior of slurries in a spiral separator becomes imperative. This necessitates the development of a comprehensive model that considers fluid flow dynamics within the helical geometry, accounting for both the free surface of the fluid and the intricate interactions between particles and the fluid. For the single phase Newtonian fluid flow (i.e., excluding particles), extensive studies have been conducted in helically curved pipes.\cite{berger1983flow,ito1987flow,germano1982effect,wang1981low} On the other hand, the fluid flow in an open helical channel presents challenges, particularly in determining the free surface. Computational fluid dynamics simulations based on Eulerian multi-fluid volume of fluid models have been used to study the separation of water, quartz, and hematite particles down a spiral separator.\cite{meng2023particulate,sudikondala2022cfd} Instead our approach investigates the fluid flows in the thin-film limit using the lubrication theory. This simplified approach build offs of the theoretical framework provided here. \cite{stokes2013thin,arnold2015thin,arnold2017thin}

Many research works \cite{zhou2005theory,cook2008theory,murisic2013dynamics,wong2019fast, mirzaeian2020bidensity} have demonstrated that, even at relatively modest volume fractions of particles, the presence of these particles significantly influences the flow dynamics, leading to distinct behaviors based on the volume fraction. Consequently, it is imperative to quantitatively model the interaction between the fluid and particles. Such models fall into two broad categories: discrete and continuum. Discrete models, which track individual particles, are relatively accurate but computationally intensive.\cite{revay1992numerical} On the other hand, continuum models treat fluid and particles as separate continuous phases, each governed by coupled equations. Our study consists of significant particle concentrations, making continuum models the preferred approach.

Two continuum models extensively explored in the realm of suspension flows are the diffusive-flux model\cite{leighton1987shear, kanehl2015hydrodynamic} and the suspension-balance model.\cite{nott1994pressure, howard2022bidisperse} The diffusive-flux model postulates a Newtonian fluid with viscosity contingent upon the particle volume fraction, while the particle flux expression is empirical. Conversely, the suspension-balance model hinges on a non-Newtonian bulk stress featuring induced normal stresses due to shear. Particle migration results from gradients in the normal stress. Consequently, the suspension-balance approach incorporates viscously generated normal stresses, playing a pivotal role in their technique. This stands in contrast to the diffusive-flux approach, where such stresses are omitted.

Previous studies employ the diffusive-flux model to investigate particle-laden thin-films moving down straight inclined planes.\cite{zhou2005theory,murisic2013dynamics,lee2015equilibrium} Their theoretical analyses align with experimental findings to reveal three distinct regimes depending on the channel slope, particle density, and initial volume fraction. In these regimes, particles either settle to the fluid bottom (settled), aggregate at the front of the slurry down the incline (ridged), or remain evenly dispersed throughout the channel depth (well-mixed). In a prior study,\cite{lee2014behavior} the diffusive-flux model was utilized to explore monodisperse particle-laden flows in helical channels with small slopes and flat channel bottom, yielding solutions in the well-mixed regime. Building upon these findings, the model's application was extended to helical channels with arbitrary centerline curvature and torsion, as well as arbitrary cross-sectional shapes such as increasing the channel bottom height in the radial direction.\cite{arnold2019particle} All these studies assume fluid layer thickness is much smaller compared with the channel width.

The aforementioned studies also assume that particles in the fluid have uniform shapes and densities. However, real-world slurries often consist of diverse types of particles because in practice, spiral separators are used to segregate distinct particle species in the slurry such as minerals from silica sand\cite{giffardpatent} or valuable coal pieces from coal ash.\cite{WrightpatentEU} By taking into account multiple species, we can optimize both the initial design and the regular adjustments of sorting equipment (such as the splitters), leading to more efficient and sustainable industrial practices. 

To address the existing knowledge gap, our work extends the current modeling framework for monodisperse particle-laden flow. We characterize the motion of the bulk particle-laden flow, comprising two particle species with identical shapes but differing densities, using a diffusive-flux model. In addition to shear-induced migration, this model considers the particle flux induced by the interaction between two different particle species. We employ an asymptotic analysis method to explore the distribution of both fluid and particles when the thickness of the fluid layer is significantly smaller than the width of the channel, hereafter called the thin-film limit. Our focus is on the flow profile near the outlet of the channel, where the flow has reached a steady state.

Previous studies\cite{lee2014behavior,arnold2019particle} show stable slurries stratify radially within the spiral separator, forming distinct zones: a concentrated inner region near the axis and a clear outer fluid zone. Our extension of this analysis of the approximated equation in the thin-film limit to multiple species generalizes the results from prior work. In bidensity particle-laden flow, the entire cross-section of the spiral separator can be discretized into three non-overlapping regions, indicating that the solution is not discontinuous in the leading-order approximation in the asymptotic analysis. The innermost region, adjacent to the inner side wall, predominantly comprises heavy particles. The middle region accommodates light particles, while the outermost region, in close proximity to the outer side wall, is primarily occupied by fluid. This analysis demonstrates that the spiral separator can separate different particle species in slurries based on their density and provides a quantitative method for determining the slurry cross-section profile in the equilibrium state. 

This paper is structured as follows: In Section \ref{sec:problem formulation}, we begin by introducing the governing equations for the particle-laden flow. Following this, we articulate the equations in the helical coordinates along with the relevant formulas with some details left in Appendix \ref{sec: app Helical coordinates system}. Subsequently, we outline the nondimensionalization procedure and present the governing equation in the thin-film limit. In Section \ref{Sec:Analysis}, we initially illustrate the particle separation in the bidensity particle-laden flow at equilibrium and subsequently derive the solution for the flow and free surface profile. In Secion \ref{sec:steadystate}, we take a look at the complete steady state solution for one practical and one fictitious apparatus. In Section \ref{sec:conclusion}, we succinctly summarize our findings and suggest potential avenues for future research.

\section{Problem Formulation}\label{sec:problem formulation}
The particles in the fluid are modeled as a continuum and are characterized by a local volume fraction, $\phi$.  We denote  $\phi_{1}$ and $\phi_{2}$ as the volume fraction of the first and second particle species, respectively. The total local volume fraction is then defined as $\phi=\phi_{1}+\phi_{2}$. Based on the diffusive flux phenomenology,\cite{murisic2013dynamics,wang2014shock,lee2015equilibrium,wong2016conservation} the particle-fluid mixture is modeled as an incompressible Newtonian fluid, with the density and viscosity modified by the presence of particles. We consider a mixture of a fluid with large viscosity $\mu_{\ell}$ and density $\rho_{\ell}$ and two species of negatively buoyant particles with diameter $d$ and densities $\rho_{1}$, $\rho_{2}$ satisfying $\rho_{\ell}\leq \rho_{1}\leq \rho_{2}$. The momentum conservation of the fluid-particle mixture is 
\begin{eqnarray} \label{eq:NS equation}
\rho (\phi) (\partial_{t} \mathbf{u} &+& \mathbf{u}\cdot \nabla \mathbf{u}) \nonumber \\
&=& - \nabla p + \nabla \cdot ( \mu (\phi)  (\nabla \mathbf{u} + \nabla \mathbf{u}^{T}) )  + \rho (\phi)\mathbf{g}. 
\end{eqnarray}
Here $\mathbf{u}$ and $p$ are the velocity vector and pressure, respectively. The vector $\mathbf{g}= (0,0,-g)$, where $g$ is gravitational acceleration. The density of the mixture is given by
\begin{eqnarray}
\rho (\phi)=\sum\limits_{i=1}^{2}\rho_{i}\phi_{i}+ \rho_{\ell} (1-\phi). 
\end{eqnarray}

We use an empirical model, the Krieger-Dougherty relation,\cite{krieger1959mechanism,howard2022bidisperse} and define local effective viscosity as
\begin{eqnarray}
\mu (\phi)=\mu_{\ell} \left( 1- \frac{\phi}{\phi_m} \right)^{-2}, 
\end{eqnarray}
where $\phi_{m}$ is the maximum volume fraction.   Morris and Boulay \cite{morris1999curvilinear} take this value to be $\phi_{m} \approx 0.68$, which we adopt here. However, we note that more recent experiments for monodisperse suspensions suggest $\phi_{m} \approx 0.59$ is more relevant for non-Brownian suspensions\cite{ovarlez2006local}.

We impose the incompressibility condition for the fluid flow, as expressed by the continuity equation
\begin{eqnarray}\label{eq:continuity equation} 
\nabla \cdot \mathbf{u}=0.
\end{eqnarray}

Coupled with the momentum equations are the conservation equations for the particle volume fractions
\begin{eqnarray}\label{eq:mass conservation}
\partial_{t}\phi_{i}+\nabla \cdot  (\phi_{i}\mathbf{u}+ \mathbf{J}_{i})=0, \quad i=1,2. 
\end{eqnarray}

We adopt the particle flux formula from prior work regarding the bidensity particle laden flow,\cite{lee2015equilibrium}
\begin{eqnarray} \label{eq: J_sum}
\mathbf{J}_{i}=\mathbf{J}_{grav,i}+ \mathbf{J}_{shear,i}+\mathbf{J}_{tracer,i}.
\end{eqnarray}
The first term in the expression for $\mathbf{J}$ corresponds to sedimentation, the settling of particles on the channel bottom due to the influence of gravity. The second  terms correspond to shear-induced migration, the tendency of particles to move away from areas of high shear and high particle volume-fraction (such as near the channel bottom) toward areas of low shear and low particle volume-fraction (such as the free surface near the center of the channel). The third term accounts for mixing between particle species due to shear-induced migration.

The flux due to settling for multiple species is
\begin{eqnarray}
\mathbf{J}_{grav,i} &=& \frac{d^{2}\mathbf{g} \phi_{i}}{18 \mu_{\ell}} \left( \left(  1- \frac{\phi}{\phi_m} \right) (\rho_{i}-\rho_{\ell}) \right. \nonumber \\
 &+& \left. \left( \frac{\mu_{\ell}\Phi (\phi)}{\mu (\phi)} - \left( 1- \frac{\phi}{\phi_m}\right) \right) \sum\limits_{j=1}^{2} (\rho_{j}-\rho_{\ell})  \frac{\phi_{i}}{\phi}\right),
\end{eqnarray}
which is generalized from prior work.\cite{tripathi1999viscous,revay1992numerical} $\Phi (\phi)$ is the hindrance function. Here, we use $\Phi (\phi)= 1-\phi$,  which is suitable in the presence of shear.\cite{schaflinger1990viscous}

The shear-induced migration term is
\begin{eqnarray}
\mathbf{J}_{shear,i}=-\frac{d^2 \phi_{i}}{4}  \left( K_{c}  \nabla (\dot{\gamma} \phi)+\frac{K_{v} \dot{\gamma} \phi}{\mu (\phi)}  \nabla \mu (\phi) \right),
\end{eqnarray}
where $\dot{\gamma}=\lVert \frac{\nabla \mathbf{u} + \nabla \mathbf{u}^{T}}{2}  \rVert $ is the shear rate, and the constants $K_{c}\approx  0.41$ and $K_{v}\approx 0.62$ are empirically determined.\cite{phillips1992constitutive} This formulation holds true for unidirectional shear flow, with limited applicability to general flows. For our investigation, the dominant flow along the helical channel dominates the secondary flow within the channel's cross-section. Consequently, we approximate the shear rate formula to the leading order, capturing the primary dynamics of interest while recognizing its restriction to situations characterized by a predominant unidirectional shear flow.

Lastly,  we include a tracer flux term  to describe the mixing between particle species due to shear-induced migration,
\begin{eqnarray}
\mathbf{J}_{tracer,i}= - \frac{\dot{\gamma} d^2}{4} D_{tr} (\phi) \phi \nabla \left( \frac{\phi_{i}}{\phi} \right).
\end{eqnarray}
Tripathi and Acrivos\cite{tripathi1999viscous} show $D_{tr}$ scales as $\phi^{2}$ for small particle volume fraction. For large concentrations, numerical simulations and experiments\cite{sierou2004shear} suggest that the tracer diffusivity becomes constant beyond a value, around $\phi=0.3\sim 0.5$. Therefore, we use the expression: $D_{tr} (\phi)=0.35\min (\phi^{2}, 0.4^{2})$. Notably, tracer diffusion drops out of the total particle flux (i.e. $\mathbf{J}_{tracer,1}+\mathbf{J}_{tracer,2}=0$), justifying its neglect in modeling monodisperse slurries on an inclined plane.\cite{murisic2013dynamics,murisic2011particle} 
Furthermore, previous studies on thin-film helical channels\cite{lee2014behavior,arnold2019particle} did not include this flux term.

At the channel walls and bottom, the velocity must vanish to satisfy the no-slip condition: $\mathbf{u}=\mathbf{0}$. At the free surface, assuming surface tension is negligible, the no stress boundary condition can be expressed as $\mathbf{n}\cdot (-p \mathbb{I}+ \mu (\phi) (\nabla \mathbf{u}+ \nabla \mathbf{u}^{T}))=\mathbf{0}$, where $\mathbb{I}$ is the identity matrix. The free surface is also governed by the kinematic condition: $\mathbf{u}\cdot \mathbf{n}=0$. The particle volume fraction satisfies no-flux boundary condition at each boundaries, namely, $\left( \phi_{i}\mathbf{u}+\mathbf{J}_{i} \right)\cdot \mathbf{n}=0$.

\subsection{Helical coordinate system}
\begin{figure}
  \centering
    \includegraphics[width=\linewidth]{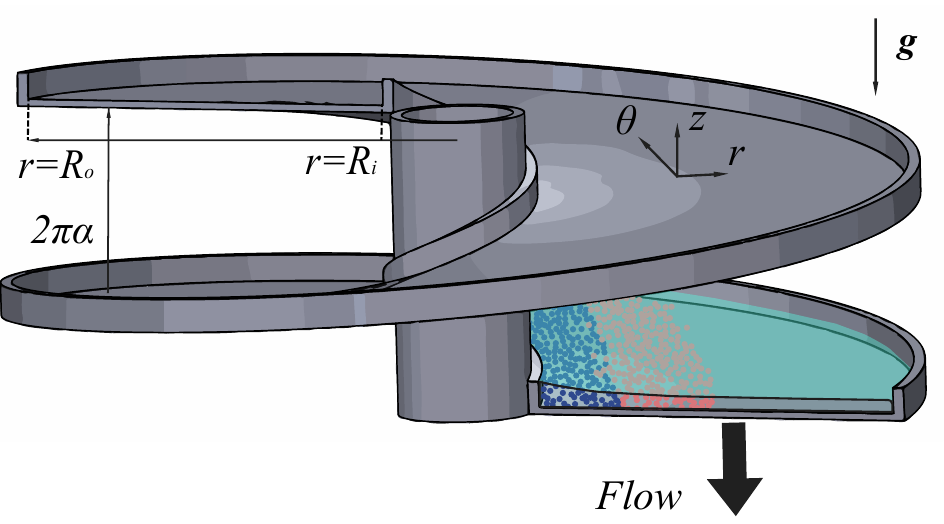}
  \hfill
  \caption[]
  { A schematic of a helical channel with rectangular cross-section.  The spiral inner radius is $R_{i}$, the outer radius is $R_{o}$, while the vertical spacing between each turn is the pitch $2\pi \alpha$. The channel width is $R=R_{o}-R_{i}$.}
  \label{fig:diagram}
\end{figure}

We describe the spiral by introducing a new helical coordinate system $(r,\theta,z)$ which is related to the position in Cartesian coordinates $\mathbf{x}(x,y,z)$ via
\begin{eqnarray} \label{eq: position}
\mathbf{x}(r,\theta,z)=r \cos \theta \mathbf{e}_{x}+ r \sin \theta \mathbf{e}_{y}+(\alpha \theta+z)\mathbf{e}_{z}. 
\end{eqnarray} 
where $r$ represents radial distance from the modiolar axis (the Cartesian $z$-axis), $\theta$ is the angle around the modiolar axis, and $z$ corresponds to the distance measured vertically from the channel floor, see Figure \ref{fig:diagram}.  The parameter $\theta$ determines how much the spiral rises per radian, so that $2\pi \alpha$ is equal to the pitch of the helix.  When $\alpha=0$, this coordinate system reduces to cylindrical coordinates. This coordinate system has applications in the modeling of fluid problems in a domain with spiral geometries.\cite{manoussaki2000effects,lee2014behavior,arnold2019particle,hill2018note}

The bounding walls are defined as follows:
\begin{eqnarray} \label{eq: walls}
\left\{ \mathbf{x} (r,\theta,z)\ |\ r\in \left\{ R_{i},R_{o} \right\},\ \theta \in \mathbb{R},\ z\in [0, 2\pi \alpha] \right\}.
\end{eqnarray}
The bottom of the channel is defined as
\begin{eqnarray} \label{eq: bottom}
\left\{ \mathbf{x} (r,\theta,z)\ |\ r\in [R_{i},R_{o}],\ \theta \in \mathbb{R},\ z=0 \right\}.
\end{eqnarray}
For fixed $r$, $\mathbf{x}(r,\theta,z)=r \cos \theta \mathbf{e}_{x}+ r \sin \theta \mathbf{e}_{y}+(\alpha \theta+z)\mathbf{e}_{z}$ is a circular helix curve. Its curvature is $r/(r^{2}+\alpha^{2})$ and the torsion is $\alpha/(r^{2}+\alpha^{2})$.

The basis vectors in the helical coordinate system are
\begin{equation} \label{eq: basis vectors cart}
\begin{cases}
\begin{aligned}
\mathbf{e}_{r} &= \cos \theta \mathbf{e}_{x}+ \sin \theta \mathbf{e}_{y}, \\
\mathbf{e}_{\theta} &= -r \sin \theta \mathbf{e}_{x}+ r\cos \theta  \mathbf{e}_{y}+ \alpha \mathbf{e}_{z},\\
\mathbf{e}_{z}&=\mathbf{e}_{z},
\end{aligned}
\end{cases}
\end{equation}
\begin{equation}
\begin{cases}
\begin{aligned}
\mathbf{e}_{x}&=\cos \theta \mathbf{e}_{r}- \frac{\sin \theta}{r} \mathbf{e}_{\theta}+\frac{\alpha \sin \theta }{r} \mathbf{e}_{z}, \\
\mathbf{e}_{y} &=\sin \theta \mathbf{e}_{r}+ \frac{\cos \theta}{r}  \mathbf{e}_{\theta}-\frac{\alpha \cos \theta }{r} \mathbf{e}_{z}, \\
\mathbf{e}_{z}&=\mathbf{e}_{z}.
\end{aligned}
\end{cases} 
\end{equation}

Note they are not all orthogonal or normalized. 
This is addressed in the helical vector calculations by using equation \eqref{eq:non_ortho_dot} in Appendix \ref{sec: app Helical coordinates system}. For a vector $\mathbf{u}= u_{x}\mathbf{e}_{x}+u_{y} \mathbf{e}_{y}+ u_{z}\mathbf{e}_{z}=u_{r}\mathbf{e}_{r}+u_{\theta} \mathbf{e}_{\theta}+ u_{z}\mathbf{e}_{z}$, the conversion between the coordinates are
\begin{equation} \label{eq: u_xyz}
\begin{cases}
\begin{aligned}
u_{x}&= \cos \theta u_{r}- r \sin \theta  u_{\theta}, \\
u_{y}&=\sin \theta u_{r}+ r \cos \theta u_{\theta}, \\
u_{z} &= \alpha u_{\theta}+ u_{z}, \\
\end{aligned}
\end{cases}
\end{equation}
\begin{equation}
\begin{cases}
\begin{aligned}
u_{r}&=\cos \theta u_{x}+\sin \theta u_{y},\\ 
u_{\theta}&=-\frac{\sin \theta}{r}u_{x}+\frac{\cos \theta}{r}u_{y},\\ 
u_{z}&= \frac{\alpha \sin \theta}{r}u_{x}- \frac{\alpha \cos \theta}{r}u_{y}+u_{z}.
\end{aligned}
\end{cases}
\end{equation}
Additional helical coordinates system vector calculus formulas can be found in Appendix \ref{sec: app Helical coordinates system}.

We are interested in the final configuration of fluid and particles near the channel exit and hence study the fully developed steady-state, helically symmetric flow. In other words, all  quantities are independent of the angular coordinate $\theta$ and time.   The continuity equation \eqref{eq:continuity equation} becomes
\begin{eqnarray} \label{eq:continuity equation 2}
\partial_{r}u_{r}+\frac{1}{r}u_{r}+\partial_{z}u_{z} =0.
\end{eqnarray} 

The Navier-Stokes equations \eqref{eq:NS equation} are then, in the coordinate direction $\mathbf{e}_{r}$,
\begin{align}
&\rho\left(  u_{r}\partial_{r}u_{r} + u_{z} \partial_{z}u_{r}- ru_{\theta}^2 \right) =  \nonumber\\
&-\partial_{r}p+ \mu \left( \left( 1+\frac{\alpha^{2}}{r^{2}} \right)\partial_{z}^{2}u_{r} + \frac{2\alpha }{r} \partial_{z} u_{\theta} - \partial_r \partial_z u_z \right)  \nonumber \\
&\,\qquad + 2\partial_{r}\mu \partial_{r}u_{r}+ \partial_{z}\mu  \left(  \left( 1+ \frac{\alpha^{2}}{r^{2}} \right) \partial_{z} u_{r}+\partial_{r}u_{z} \right),
\end{align}
in the coordinate direction $\mathbf{e}_{\theta}$,
\begin{align}
&\rho \left(  u_{r} \partial_{r}u_{\theta}+ u_{z} \partial_{z}u_{\theta}+ \frac{2}{r} u_{r} u_{\theta} \right)= \nonumber \\
&\frac{\alpha}{r^{2}} \partial_{z}p+ \mu  \left( \partial_{r}^{2} u_{\theta}+ \left( 1+\frac{\alpha^{2}}{r^{2}} \right) \partial_{z}^{2} u_{\theta}+ \frac{3}{r} \partial_{r} u_{\theta}- \frac{2 \alpha }{r^{3}}\partial_{z} u_{r}  \right) \nonumber \\
& \qquad + \partial_{z}\mu \left( - \frac{2\alpha }{r^{3}} u_{r}+\left( 1+\frac{\alpha^{2}}{r^{2}} \right) \partial_{z} u_{\theta}- \frac{\alpha }{r^{2}}\partial_{z} u_{z} \right),  \nonumber \\
& \qquad \qquad + \partial_{r}\mu \left( - \frac{\alpha }{r^{2}} \partial_{z} u_{r}+ \partial_{r}u_{\theta} \right)
\end{align}
and in the coordinate direction $\mathbf{e}_{z}$
\begin{align}
&\rho \left( u_{r}\partial_{r}u_{z}+ u_{z}\partial_{z}u_{z}-\frac{2\alpha}{r} u_{r}u_{\theta} \right) =- \left( 1+ \frac{\alpha^{2}}{r^{2}} \right) \partial_{z}p \nonumber\\
&+\mu \left( \partial_{r}^{2}u_{z}+ \left( 1+ \frac{\alpha^{2}}{r^{2}} \right) \partial_{z}^{2}u_{z}+ \frac{2 \alpha^{2}}{r^{3}} \partial_{z}u_{r}+ \frac{1}{r}\partial_{r}u_{z}-  \frac{2\alpha}{r} \partial_{r}u_{\theta} \right) \nonumber\\
&+ \partial_{r}\mu  \left( \left( 1+ \frac{\alpha^{2}}{r^{2}} \right) \partial_{z}u_{r}+ \partial_{r}u_{z} \right) \nonumber\\
&+ 2\partial_{z}\mu \left( \frac{\alpha^{2}}{r^{3}}u_{r}+ \left( 1+ \frac{\alpha^{2}}{r^{2}} \right)\partial_{z}u_{z} \right)- \rho g, 
\end{align}
where the continuity equation \eqref{eq:continuity equation} is used to simplify the expression of $\nabla \cdot ( \mu (\phi)  (\nabla \mathbf{u} + \nabla \mathbf{u}^{T}) )$.

Starting with Eq.~\eqref{eq:mass conservation}, we can apply the divergence (Eq.~\eqref{eq: divergence}) and the $\theta-$independent condition to derive the particle transport equation for a general particle flux vector $\mathbf{J}_{i}=J_{i,r}\mathbf{e}_{r}+J_{i,\theta}\mathbf{e}_{\theta}+J_{i,z}\mathbf{e}_{z}$ to be
\begin{align}\label{eq:particle volume fraction helical coordinates}
&\partial_{t}\phi_{i}+ u_{r} \partial_{r}\phi_{i}+u_{z}\partial_{z}\phi_{i}+\sum\limits_{i=1}^{2}\left(\frac{1}{r} \partial_{r} (rJ_{i,r})+\partial_{z}J_{i,z} \right)=0,
\end{align}
for $i=1,2$. From Eq.~\eqref{eq: J_sum}, the relevant components of the particle flux vector are
\begin{subequations}
\begin{align}
&\mathbf{J}_{i,r}= - \frac{d^{2}\phi_{i} }{4} \left(  K_{c}  \partial_{r} (\dot{\gamma} \phi)+ \frac{K_{v} \dot{\gamma} \phi}{\mu (\phi)}  \partial_{r} \mu (\phi)  \right) \nonumber \\
&\qquad - \frac{\dot{\gamma} d^2}{4} D_{tr} (\phi) \phi \partial_{r} \left( \frac{\phi_{i}}{\phi} \right) ,\\
&  \mathbf{J}_{i,z}=- \frac{d^{2}\phi_{i}\left( 1+\frac{\alpha^{2}}{r^{2}} \right) }{4} \left( K_{c}  \partial_{z} (\dot{\gamma} \phi)+  \frac{K_{v} \dot{\gamma} \phi}{\mu (\phi)} \partial_{z} \mu (\phi) \right) \nonumber \\
& \qquad -\frac{d^{2}g \phi_{i}}{18 \mu_{\ell}} \left( \left(  1- \frac{\phi}{\phi_m} \right) (\rho_{i}-\rho_{\ell}) \right. \nonumber \\
 & \qquad + \left. \left( \frac{\mu_{\ell}\Phi (\phi)}{\mu (\phi)} - \left( 1- \frac{\phi}{\phi_m}\right) \right) \sum\limits_{j=1}^{2} (\rho_{j}-\rho_{\ell})  \frac{\phi_{i}}{\phi}\right) \nonumber \\
& \qquad \qquad - \frac{\dot{\gamma} d^2}{4} D_{tr} (\phi) \phi  \left( 1+\frac{\alpha^{2}}{r^{2}} \right)\partial_{z}  \left( \frac{\phi_{i}}{\phi} \right).
\end{align}
\end{subequations}
 We omit the expression of $J_{i,\theta}$ since it will not be used in the governing equations.

We denote the free surface as $z=h (r)$. The kinematic boundary condition $\mathbf{n} \cdot \mathbf{u} = 0$ at the free surface becomes
\begin{eqnarray}
&u_{z}-u_{r}\partial_{r}h=0.
\end{eqnarray}
The no stress boundary condition at the free surface $\mathbf{n} \cdot (-p \mathbb{I} +\mu (\phi)  (\nabla \mathbf{u} + \nabla \mathbf{u}^{T}) ) = \mathbf{0}$ becomes
\begin{subequations}
\begin{align}
  &\mu \left(-2\partial_{r}h \partial_{r}u_{r}+\partial_{r}u_{z}+\left( 1+ \frac{\alpha^{2}}{r^{2}} \right)\partial_{z}u_{r}\right)+p\partial_{r}h=0, \\
  &\mu \left(-\partial_{r}h \left(\partial_{r}\partial  u_{\theta}-\frac{\alpha}{r^{2}} \partial_{z}u_{r}\right)+\left( \left( 1+\frac{\alpha^2}{r^{2}} \right)\partial_{z}u_{\theta}- \frac{\alpha}{r^{2}} \partial_{z}u_{z}\right)\right. \nonumber \\
  &\qquad \qquad  \qquad  \qquad \qquad \qquad  \left.-\frac{2\alpha}{r^{3}}  u_{r}\right)+\frac{\alpha p}{r^2}=0, \\
  &\mu \left(- \partial_{r}h \left( \partial_{r}u_{z}+\left( 1+\frac{\alpha ^2}{r^{2}} \right)\partial_{z}u_{r}\right)+2 \left( 1+\frac{\alpha ^2}{r^{2}} \right)\partial_{z}u_{z}\right. \nonumber \\
  &\qquad \qquad  \qquad  \qquad \left.+\frac{2\alpha ^2}{r^{3}} u_{r}\right)-\left( 1+\frac{\alpha^2}{r^2} \right)p=0.
\end{align}
\end{subequations}
The no-flux boundary condition for the particle volume fraction is
\begin{eqnarray}
(u_{z}\phi_{i}+\mathbf{J}_{z,i})-(u_{r}\phi_{i}+\mathbf{J}_{r,i})\partial_{r}h=0,\quad i=1,2.
\end{eqnarray}

\subsection{Nondimensionalization and thin-film approximation} \label{sec:Nondimensionalization and thin-film approximation}
The fluid flow in the spiral separator is driven by gravity. Therefore, the velocity component along the spiral channel direction is much larger than the velocity component in other directions, namely $ru_{\theta}\gg u_{r}, u_{z}$. Typically, the depth of fluid layer $H$ is small relative to the channel width $R=R_{o}-R_{i}$, namely, $\epsilon= H/R \ll 1$. Then to balance the terms in \eqref{eq:continuity equation 2}, we have $u_{r}\gg u_{z}$.
Notice that in the helical coordinate system, $\mathbf{e}_{\theta}$ is not normalized to be a unit vector. Therefore,  $u_{\theta}$ is the angular velocity (with units of $1/t$).

To nondimensionalize we use the following change of variables for the thin-film approximation
\begin{align} \label{eq:nondimensionalization}
  &Rr'=r , \quad Hz'=z, \quad \epsilon Uu_{r}'=u_{r}, \quad  \frac{U}{R} u_{\theta}'=u_{\theta},\quad \epsilon^{2} U u_{z}'=u_{z}, \nonumber \\
  & \frac{\mu_{\ell}U}{H} p'=p,\quad \rho_{\ell}\rho'=\rho,\quad \mu_{\ell}\mu'=\mu. 
\end{align}
We also introduce the dimensionless parameters:  $\mathrm{Re}=\rho_{\ell}U H/\mu_{\ell}$ is the Reynolds number and $\mathrm{Ri}= gH/U^{2}$ is the Richardson number. After nondimensionalization, the range of $r$ is $R_{i}'=R_{i}/R\leq r\leq R_{o}/R=R_{o}'$. 
The crucial element in this analysis is the scaling of the particle transport equation describing the fast equilibration of particles in the direction normal to the subtrate \cite{murisic2013dynamics}. As in the incline problem, we require that $ \left( d/H \right)^{2} \ll 1$, ensuring that the particle size is consistent with the continuum model. Additionally, we assume ($\epsilon \ll (d/H)^{2}$)  to prevent the system from reaching a colloidal limit, where the particles are too small and Brownian motion becomes dominant. This assumption also makes sure the particle reaches equilibrium distribution at the time scale investigated in this work. If we denote $\left( d/H \right)^{2}=\epsilon^{\beta}$, then $0<\beta<1$. Therefore, we can nondimensionalize the particle flux term as 
\begin{align}
&\left( \frac{d}{H} \right)^{2} \frac{H}{R}U J_{i,r}'= \epsilon^{\beta+1}U J_{i,r}'=J_{i,r},\nonumber \\
&\left( \frac{d}{H} \right)^{2} U J_{i,z}'= \epsilon^{\beta}U J_{i,z}'=J_{i,z},\quad i=1,2.
\end{align}
The prime over the dimensionless quantities will be subsequently dropped for brevity. Then the continuity equation \eqref{eq:continuity equation 2} and the conversation of the volume fraction equation \eqref{eq:particle volume fraction helical coordinates} become
\begin{subequations}
\begin{align}
  &\frac{1}{r}\partial_{r} (r u_{r})+ \partial_{z}u_{z}=0,  \\
  & \epsilon^{2}( u_{r} \partial_{r}\phi+ u_{z}\partial_{z}\phi)+\sum\limits_{i=1}^{2}\left( \epsilon^{\beta+1}   \frac{1}{r} \partial_{r}(rJ_{i,r})+\epsilon^{\beta}\partial_{z}J_{i,z} \right)=0. \label{eq:volume fraction nondimensionalization}
\end{align}
\end{subequations}

The Navier-Stokes equation takes the form of equations \eqref{eq: full NS} seen in Appendix \ref{sec: app Helical coordinates system}. Let $\epsilon\rightarrow 0$ and fix other parameters,  the leading order approximations for the above governing equations are
\begin{subequations}
  \begin{align}
    &\frac{1}{r}\partial_{r} (r u_{r})+ \partial_{z}u_{z}=0, \label{eq:continuity nondimensionalization} \\
 &\partial_{r}p- \left(1+\frac{\alpha^{2}}{r^2 R^2}  \right)\partial_{z} (\mu \partial_{z}u_{r}) -\frac{2\alpha \mu  }{r R}\partial_{z}u_{\theta}-r\text{Re}\rho   u_{\theta} ^2=0, \label{eq:leading order equation b}\\
&- \frac{\alpha}{r^{2}R} \partial_{z}p- \left(1+\frac{\alpha^{2}}{r^2 R^2}  \right)\partial_{z} (\mu \partial_{z}u_{\theta}) =0,  \label{eq:leading order equation c} \\
  &\left(1+\frac{\alpha^{2}}{r^2 R^2}  \right)\partial_{z}p+\text{Re}\text{Ri}\rho  =0. \label{eq:leading order equation d}
\end{align}
\end{subequations}

Similar to previous studies,\cite{murisic2013dynamics,wang2014shock,lee2015equilibrium,wong2016conservation} we deduce that, at the leading order, the conservation of the particle volume fraction \eqref{eq:volume fraction nondimensionalization} implies
\begin{equation}
\begin{aligned}
&\partial_{z}\mathbf{J}_{i,z}=0, \quad i=1,2.\
\end{aligned}
\end{equation}
Moreover, additional insights can be derived from equation \eqref{eq:volume fraction nondimensionalization}. Given that $0<\beta<1$,
\begin{eqnarray}
& u_{r}\partial_{r}\phi_{i}+ u_{z}\partial_{z}\phi_{i}=0,  \quad i=1,2. \label{eq: leading-order constraint}
\end{eqnarray} 
In previous spiral channel models,\cite{lee2014behavior,arnold2019particle} a mass conservation condition is imposed alongside the conservation of the particle volume fraction, leading to an equation equivalent to the one above at the leading-order approximation.  The observations we make here contribute to a framework that aligns with previous models \cite{lee2014behavior,arnold2019particle} without introducing additional conservation constraints.

The leading order approximation for the boundary condition at the free surface is
\begin{subequations}
\begin{align}
&p \partial_{r}h+ \mu\left(1+\frac{\alpha^{2}}{r^2 R^2}  \right)\partial_{z}u_{r}=0, \label{eq: bc leading-order u_r}\\
&\frac{\mu}{R}\left(1+\frac{\alpha^{2}}{r^2 R^2}  \right)\partial_{z}u_{\theta}+\frac{\alpha p }{r^2 R^2}=0, \label{eq: bc leading-order u_thet}\\
&-\mu\left(1+\frac{\alpha^{2}}{r^2 R^2}  \right)p =0. \label{eq: bc leading-order pressure}
\end{align}
\end{subequations}
Equation \eqref{eq: bc leading-order pressure} implies $p=0$ at $z=h(r)$. With this conclusion, equations \eqref{eq: bc leading-order u_r} and \eqref{eq: bc leading-order u_thet} imply $\partial_{z}u_{r}=\partial_{z}u_{\theta}=0$ at $z=h (r)$.

The kinematic boundary condition remains the same: 
\begin{eqnarray}
u_{z}-u_{r}\partial_{r}h=0. \label{eq: kinematic BC}
\end{eqnarray}

The boundary condition for the particle flux at the free surface becomes $\mathbf{J}_{z,i}=0$. The asymptotic expansion of the shear rate is $\dot{\gamma} = r\left(1+\alpha^{2}/(r^2 R^2)  \right) |\partial_{z}u_{\theta}|+\mathcal{O} (\epsilon)$. Therefore, the leading order approximation of particle flux in $z$-direction is 
\begin{align}\label{eq:particle flux leading order}
  &  \mathbf{J}_{i,z}= - \frac{r |\partial_{z}u_{\theta}|}{4} D_{tr} (\phi) \phi  \left( 1+\frac{\alpha^{2}}{R^{2}r^{2}} \right)^{2}\partial_{z}  \left( \frac{\phi_{i}}{\phi} \right) \nonumber\\
  &-\frac{r \phi_{i}}{4}   \left( 1+\frac{\alpha^{2}}{R^{2}r^{2}} \right)^{2} \left( K_{c}  \partial_{z} (\phi|\partial_{z}u_{\theta}|) +\frac{K_{v} \phi}{\mu (\phi)} |\partial_{z}u_{\theta}|\partial_{z} \mu (\phi)  \right) \nonumber \\
  &-\frac{\mathrm{Re} \mathrm{Ri}\phi_{i}}{18 } \left(  \left(  1- \frac{\phi}{\phi_m} \right) (\rho_{i}-1) \right. \nonumber \\
  &\qquad \qquad \quad \left.+ \left( \frac{\Phi (\phi)}{\mu (\phi)} - ( 1- \frac{\phi}{\phi_m}) \right) \sum\limits_{j=1}^{2} (\rho_{j}-1)  \frac{\phi_{i}}{\phi}\right).
\end{align}

Integrating $\partial_{z}\mathbf{J}_{z,i}=0$ once yields $\mathbf{J}_{z,i}=C$, where $C$ is a constant number. The boundary condition of the particle flux at the free surface implies $C=0$. Therefore, we obtain $\mathbf{J}_{z,i} (r,z) =0$ for any $r,z$ in the domain.

We derive the following two integral forms of the continuity equation using the free surface conditions. 
The first form is solved as follows:
\begin{subequations}
\begin{align}
\partial_{r} \int\limits_{0}^{h (r)} r u_{r}\mathrm{d} z&=\left. r u_{r} \right|_{z=h (r) }\partial_{r}h+  \int\limits_{0}^{h (r)} \partial_{r} (r u_{r})\mathrm{d}z \\
& =\left. r u_{r} \right|_{z=h (r) }\partial_{r}h-  \int\limits_{0}^{h (r)} r\partial_{z}u_{z}\mathrm{d}z \label{eq: h(r) step 2} \\
& =r \left( \left.  u_{r}\partial_{r}h -u_{z} \right|_{z=h (r) }\right)=0. \label{eq: h(r) step 3}
\end{align}
\end{subequations}
Equation \eqref{eq: h(r) step 2} follows the continuity equation. Equation \eqref{eq: h(r) step 3} follows the kinematic boundary condition at the free surface. 
This result implies $\int_{0}^{h (r)} r u_{r}\mathrm{d}z$ is a constant. Notice that due to the no-slip boundary condition, $u_{r}=0$ at the side wall (where $r>0$).    Therefore, in order to be constant it must be the case $\int_{0}^{h (r)}  ru_{r}\mathrm{d}z=0$ for all $r$. A further simplification would be to remove $r$ since the integral is respect to $z$. This aligns with the integral form of the continuity equation presented in previous work.\cite{lee2014behavior, arnold2019particle}
We next derive the other integral form with a similar procedure
\begin{subequations}
\begin{align}
\partial_{r} \int\limits_{0}^{h (r)} r u_{r}\phi_{i}\mathrm{d} z&=\left. r u_{r}\phi_{i} \right|_{z=h (r) }\partial_{r}h+  \int\limits_{0}^{h (r)} \partial_{r} (r u_{r}\phi_{i})\mathrm{d}z \\
& =\left. r u_{r}\phi_{i} \right|_{z=h (r) }\partial_{r}h-  \int\limits_{0}^{h (r)} r\partial_{z}\left( u_{z}\phi_{i} \right)\mathrm{d}z \\
& = \left. r \phi_{i}\left( u_{r}\partial_{r}h-u_{z} \right) \right|_{z=h (r) } =0. 
\end{align}
\end{subequations}

Further investigation of the particle volume fractions requires solving $\mathbf{J}_{i,z}=0$, where $\mathbf{J}_{z,i}$ is provided in equation \eqref{eq:particle flux leading order}. Multiply by $4\mu /\phi_{i}$ to rewrite it as follows:
\begin{align}
  0=&- r\mu|\partial_{z}u_{\theta}|D_{tr} (\phi) \frac{\phi}{\phi_{i}} \left( 1+\frac{\alpha^{2}}{R^{2}r^{2}} \right)^{2}\partial_{z}  \left( \frac{\phi_{i}}{\phi} \right) \nonumber \\
  &-r\left( 1+\frac{\alpha^{2}}{R^{2}r^{2}} \right)^{2} \left( K_{c}  \mu \partial_{z} (\phi|\partial_{z}u_{\theta}|) +K_{v} \phi |\partial_{z}u_{\theta}|\partial_{z} \mu \right)\nonumber \\
  &-\frac{2\mu \mathrm{Re} \mathrm{Ri}}{9 } \left(  \left(  1- \frac{\phi}{\phi_m} \right) (\rho_{i}-1) \right. \nonumber \\
  &+\left.\left( \frac{\Phi (\phi)}{\mu (\phi)} - ( 1- \frac{\phi}{\phi_m}) \right) \sum\limits_{j=1}^{2} (\rho_{j}-1)  \frac{\phi_{i}}{\phi}\right), \quad i = 1,2.
\end{align}
From equation \eqref{eq:continuity nondimensionalization}, one can show $\partial_{z}u_{\theta}\leq 0$ for $0\leq z\leq h (r)$. Therefore,  we remove the absolute value function in above equation. Second, defining shear stress $\sigma (r,z) = r\left(1+\frac{\alpha^{2}}{r^2 R^2}  \right)^{2}\mu \partial_{z}u_{\theta}$, we have the following system of equations
\begin{subequations}
\begin{align}
  & 0=  \sigma D_{tr} (\phi)  \partial_{z} \ln \left( \frac{\phi_{i}}{\phi} \right)+ K_{c}   \partial_{z} (\phi\sigma) +(K_{v}-K_{c}) \phi  \sigma  \partial_{z} \ln(\mu) \nonumber \\
  &-\frac{2\mu \mathrm{Re} \mathrm{Ri}}{9 }  \left(  \left(  1- \frac{\phi}{\phi_m} \right) (\rho_{i}-1)+ \right. \label{eq:equilibrium flux 2a} \\
  &\qquad \qquad \quad \left. \left( \frac{\Phi (\phi)}{\mu (\phi)} - \left( 1- \frac{\phi}{\phi_m}\right) \right) \sum\limits_{j=1}^{2} (\rho_{j}-1)  \frac{\phi_{i}}{\phi}\right),\nonumber \\
 & \partial_{z} \sigma =\frac{\alpha \text{Re}\text{Ri} \rho}{ rR }, \label{eq:equilibrium flux 2b} \\
 &\left.  \sigma \right|_{z=h (r) }=0. \label{eq:equilibrium flux 2c}
\end{align}
\end{subequations}
An additional constraint is required to solve this equation: the integral of the volume fraction equals to a given constant $\int\limits_0^{h}\phi \mathrm{d}z= C$.

\section{Analysis}
\label{Sec:Analysis}
\subsection{Particle distribution }
For a fluid mixed with a single particle species, established models\cite{lee2014behavior,arnold2019particle} demonstrate that the stable configuration of the particle distribution at the steady state within a spiral channel cross-section comprises two distinct regions. The outer region, adjacent to the outer wall, is characterized by the presence of clear fluid exclusively. The inner region, closest to the spiral's center axis, holds particles with a concentration that shows no vertical variation.

\begin{figure}
  \centering
    \includegraphics[width=\linewidth]{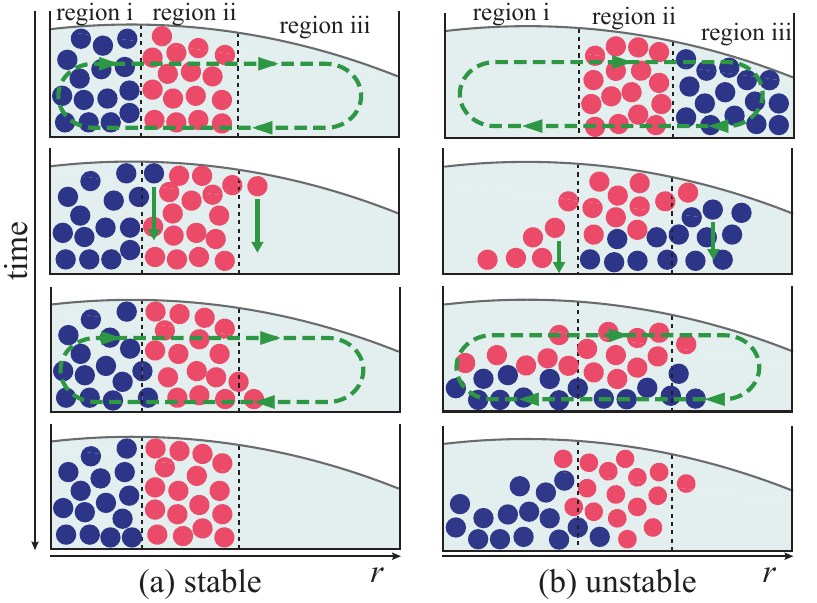}
  \hfill
  \caption[]
  {The stable (a) and unstable (b) equilibrium configuration. Heavy particles are shown in dark blue and lighter particles are red. The rows increase in time from top to bottom, demonstrating the behavior of different starting configurations when perturbed by the secondary flow.}
  \label{fig:stability}
\end{figure}

Expanding upon this concept, we postulate the existence of three distinct regions in the bidensity particle-laden flow, as illustrated in Fig.~\ref{fig:stability}. In region (\romannumeral 1), $R_{i}<r<R_{p}$, predominantly heavier particles are found. Region (\romannumeral 2), $R_{p}<r<R_{f}$, harbors lighter particles. Region (\romannumeral 3), $R_{f}<r<R_{o}$, is exclusively clear liquid. In all regions, the volume fractions for each particle species remain independent of the vertical coordinates. In fact, this type of particle distribution is the only one that fulfills the constraint $\int_0^{h (r)}\phi_{i}u_{r}\mathrm{d} z=0$ when the velocity satisfies $\int_0^{h (r)}u_{r}\mathrm{d}z=0$.

In order to find the particle distributions that are independent of $z$, we can set $\partial_{z}\phi_{1}=0$, $\phi_{2}=0$ (or  $\partial_{z}\phi_{2}=0$, $\phi_{1}=0$) in equation \eqref{eq:equilibrium flux 2a}, 
\begin{align}\label{eq:critical volume fraction equation}
0=  \phi_{i}^{2}+\phi_{i}\left( \frac{2 r R}{9 \alpha  K_c}+\frac{1}{\rho _1-1} \right)-\frac{2 r R}{9 \alpha  K_c}, \quad i=1,2.
\end{align}
Recall that $0\leq\phi\leq\phi_{m}$. Then we obtain the formula for the critical particle volume fraction
\begin{align}\label{eq:critical volume fraction}
\phi_{c,i}(r)=&\min \left\{ \phi_{m}, \frac{1}{2} \left(\sqrt{\left(\frac{2 r R}{9 \alpha  K_c}+\frac{1}{\rho _1-1}\right)^2+\frac{8 r R}{9 \alpha  K_c}} \right. \right. \nonumber \\
&\left. \left. -\left( \frac{2 r R}{9 \alpha  K_c}+\frac{1}{\rho _1-1} \right)\right) \right\}. 
\end{align}
Now we can describe the particle volume fraction in each region of the stable configuration more precisely.  In region  (\romannumeral 1), $\phi_{1} (r,z) =0$ and $\phi_{2} (r,z) =\phi_{c,2} (r)$.  In region  (\romannumeral 2), $\phi_{1} (r,z) =\phi_{c,1} (r)$ and $\phi_{2} (r,z) =0$.  In region  (\romannumeral 3), $\phi_{1} (r,z) =0$ and $\phi_{2} (r,z) =0$.

\subsection{Stability analysis}
To justify the proposed configuration, we need to demonstrate the interface between each region is sharp, and the configuration is stable.

\subsubsection{Interface thickness}
First, we will analyze the particle dynamics in the vicinity of each interface, demonstrating that the interface thickness is zero in the thin-film limit.  Viewing the channel cross-section with the spiral's axis on the left, the clockwise secondary flow transports particles from one region to another (see Fig.~\ref{fig:stability}), and the particles rapidly establish an equilibrium profile in the vertical direction due to the thin fluid layer thickness. Specifically, the vertical particle flux term scales as $\partial_{z}\mathbf{J}_{i}\sim \epsilon^{\beta} U/H$. To maintain balance with this term, the corresponding timescale should be $(H/U)\epsilon^{-\beta}=(R/U)\epsilon^{1-\beta}$.  It is important to note that the timescale for the thin-film approximation, discussed in Section \ref{sec:Nondimensionalization and thin-film approximation}, is the bulk flow timescale, $(R/U)$, which is larger than the timescale for reaching equilibrium in the vertical direction, given that $0<\beta<1$.  In other words, we can reasonably assume the perturbed particles immediately return to equilibrium at the bulk flow timescale we considered in the thin film approximation.

The thickness of the interface between each region is determined by the length scale of the region where particles undergo transient dynamics. Estimating this length scale involves multiplying the timescale of the transient dynamics by the radial velocity scale, yielding $(H/U)\epsilon^{-\beta} \epsilon U= H \epsilon^{1-\beta}=R\epsilon^{2-\beta}$. As $\epsilon \rightarrow 0$, the length scale of the transient region is significantly smaller than the channel width and is even smaller than the characteristic fluid depth $H=R \epsilon$.  Consequently, we assume the interface thickness is zero, and our primary focus is on the particle equilibrium profile rather than delving into transient dynamics. In addition, this observation implies the particle distribution and fluid flow may not be continuous within the cross-section of the channel.
Having established the sharpness of the interface, the subsequent two subsections concentrate on demonstrating the stability of the configuration through particle interaction and the clockwise secondary flow.

\subsubsection{Interface between pure fluid and single species slurry}
Consider the interface between region (\romannumeral 2) and (\romannumeral 3) in the stable configuration.  In this scenario, the problem simplifies to a particle-laden flow with a single species, akin to the previous studies.\cite{stokes2013thin,arnold2015thin} By setting $\phi_{2}=0$ and $\phi=\phi_{1}$ in equations \eqref{eq:equilibrium flux 2a}-\eqref{eq:equilibrium flux 2c}, we can express the systems of equations conveniently as:
\begin{subequations}
\begin{align}
&\partial_{z} \phi=\frac{  \frac{\text{Re}\text{Ri} \alpha (1-\rho_{1})}{rR} (\phi-\phi_{+c}) (\phi-\phi_{-c})}{\left( 1 + \frac{(K_{v}-K_{c})}{K_{c}}   \frac{2\phi }{\phi_{m}-\phi} \right)  \sigma  }, \label{eq:equilibrium flux region 23} \\
& \partial_{z} \sigma =\frac{\alpha \text{Re}\text{Ri}  (1+(\rho_{1}-1)\phi)}{ rR },\\
&\left. \sigma \right|_{z=0} =\frac{-\alpha \text{Re}\text{Ri} }{ r^{2}R } \left(  h (r)  +(\rho_{1}-1)\int\limits_{0}^{h (r)}\phi\mathrm{d}z  \right),\\
&\left.  \sigma \right|_{z=h (r) }=0,  \label{eq:equilibrium flux region 23d} 
\end{align}
\end{subequations}
where $\phi_{\pm c}$ is also the solution of equation \eqref{eq:critical volume fraction equation}:
\begin{align}
\phi_{\pm c} =  &\frac{1}{2} \left(\pm \sqrt{\left(\frac{2 r R}{9 \alpha  K_c}+\frac{1}{\rho _1-1}\right)^2+\frac{8 r R}{9 \alpha  K_c}}\right.\nonumber \\
&\left. - \left( \frac{2 r R}{9 \alpha  K_c}+\frac{1}{\rho _1-1} \right)\right). \label{eq:phi pmcrit}
\end{align}
From equation \eqref{eq:phi pmcrit}, we find that $\sigma (r,z) \leq 0$ for $0 \leq z \leq h(r)$ and $R_{i}<r<R_{o}$. When $\phi < \phi_{+c}$, $\partial_{z}\phi$ is negative, causing $\phi$ to decrease with increasing $z$. Conversely, when $\phi > \phi_{+c}$, $\partial_{z}\phi$ is positive, leading to a monotonic increase in $\phi$ with increasing $z$. Notably, as $z$ approaches $h(r)$, $\sigma$ tends toward $0$, and $\partial_{z}\phi$ becomes large, causing $\phi$ to rapidly reach the maximum volume fraction or decrease to zero.

\begin{figure}
  \centering
    \subfigure[ ~Dilute]{
    \includegraphics[width=0.8\linewidth]{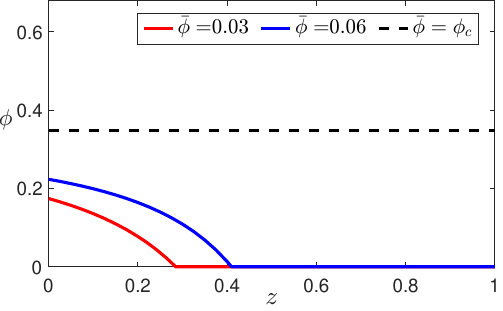}
  }
  \subfigure[ ~Critically concentrated]{
    \includegraphics[width=0.8\linewidth]{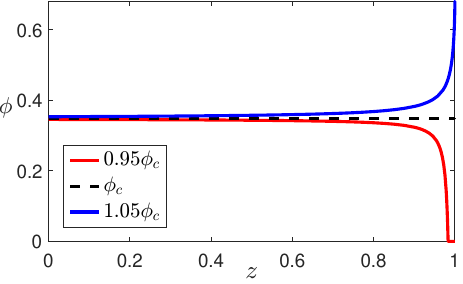}
    }
  \hfill
  \caption[]
  { The vertical equilibrium single species particle concentration profile for different $\bar{\phi}=\int_{0}^{h (r)}\phi\mathrm{d}z$ by numerically solving equations \eqref{eq:equilibrium flux region 23}-\eqref{eq:equilibrium flux region 23d}. The parameters are the  $\rho_{2}=2.5$,  $\alpha=1$, $R=1$, $r=1$, and $h=1$. Equation \ref{eq:critical volume fraction} yields $\phi_{c}= 0.3481$ which is marked with a dashed line. Panel (a) shows two dilute scenarios, when $\bar{\phi}$ is near 0, in red and blue. The particles concentrate near the channel bottom. Panel (b) shows that when $\bar{\phi} > \phi_{c}$ (blue), the extra particles are near the surface, and when $\bar{\phi} > \phi_{c}$ (red), the fluid is near the free surface.  
  }
  \label{fig:criticalVolumeFraction}
\end{figure}

Now we can analyze the particle dynamics in each regions. Region (\romannumeral 3) comprises entirely of the fluid. When a small amount of particles is introduced into this region, the particle distribution rapidly converges to the equilibrium state after the perturbation. Notably, with a small total volume of particles, they tend to concentrate near the bottom, as illustrated in panel (a) of Fig.~\ref{fig:criticalVolumeFraction}. Therefore, once the secondary flow near the free surface transports particles from the particle-rich region into the clear-fluid area,  these particles swiftly settle at the channel bottom.  From there, the particles are subsequently transported back to the particle-rich region by the secondary flow near the channel bottom, which restores equilibrium in the $r$-direction.

Region (\romannumeral 2) is characterized by an abundance of lighter particles at the critical volume fraction. The secondary flow near the channel bottom facilitates the transportation of more fluid or particles from the clear fluid region to the particle-rich region. When these additional particles are introduced to this region, we observe $\bar{\phi} > \phi_{c}$. As depicted in panel (b) of Fig.~\ref{fig:criticalVolumeFraction}, these extra particles concentrate near the free surface in the equilibrium state. Alternatively, when more fluid is added to this region, $\bar{\phi}$ becomes less than $\phi_{c}$. Panel (b) of Fig.~\ref{fig:criticalVolumeFraction} illustrates that the extra fluid occupies the space closest to the free surface in the equilibrium state as the particles will want to settle. Consequently, in both scenarios, the extra fluid or particles concentrate near the free surface and are sequentially carried to region (iii) by the secondary flow, ultimately restoring equilibrium in the $r$-direction.

\subsubsection{Interface between two particle species slurries}

\begin{figure}
  \centering
  \subfigure[ ~Mostly light particles (red)]{
    \includegraphics[width=0.8\linewidth]{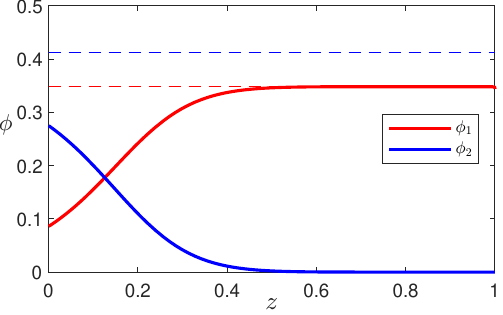}
  }
  \subfigure[ ~Mostly heavy particles (blue)]{
    \includegraphics[width=0.8\linewidth]{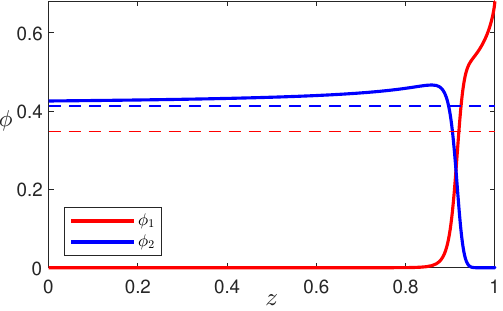}
  }
  \hfill
  \caption[]
  {The vertical equilibrium particle concentration profile for different $\bar{\phi}_{i}=\int_{0}^{h (r)}\phi_{i}\mathrm{d}z$, $i=1,2$ by numerically solving equations \eqref{eq:equilibrium flux region 12}-\eqref{eq:equilibrium flux region 12e}. The parameters are  $\rho_{1}=2.5$, $\rho_{2}=3.8$, $r=1$, $R=1$, $\alpha=1$, and $h=1$. Equation \eqref{eq:critical volume fraction} yields $\phi_{c,1}=0.3481$ and $\phi_{c,2}=0.4131$ shown as dashed lines. The red and blue dashed line represents the critical volume fraction for the light and heavy particle species respectively.  Panel (a) shows a case when small amount heavy particles ($\bar{\phi}_{2}=0.05$) is added into region  (\romannumeral 2) which mainly consists of the light particles ($\bar{\phi}_{1}=0.30$). Panel (b) shows the case when small amount of light particles ($\bar{\phi}_{1}=0.05$) is added into heavy particle concentrated region ($\bar{\phi}_{2}=0.4$).}
  \label{fig:criticalVolumeFractionBi}
\end{figure}

Next, we focus on the interface between regions (\romannumeral 1) and (\romannumeral 2), namely, the interface between slurries of two different particle species. To investigate the vertical equilibrium profile for a bidensity slurry, it is convenient to write equations \eqref{eq:equilibrium flux 2a}-\eqref{eq:equilibrium flux 2c} in the following form

\begin{subequations}
\begin{align}
  \partial_{z}\phi&= \frac{\text{Re}\text{Ri}}{\left( 1+\frac{K_{v}-K_{c}}{K_{c}} \phi  \partial_{\phi} \ln(\mu) \right)\sigma}\left( -\frac{\alpha  \rho \phi}{ r R }   \right.\nonumber \\
  &\quad \left.+\frac{2 (1-\phi) }{9 K_{c} }  \left( (\rho_{1}-1)\chi+ (\rho_{2}-1) (1-\chi) \right) \right), \label{eq:equilibrium flux region 12} \\
  \partial_{z} \chi&=\frac{2 \chi (1-\chi) \mathrm{Re} \mathrm{Ri}(\rho_{1}-\rho_{2})}{9\sigma D_{tr} (\phi) \left( 1-\frac{\phi}{\phi_{m}} \right)}, \\
  \sigma &=\int\limits_{h (r)}^{z}\frac{\alpha \text{Re}\text{Ri} \rho}{ r R }\mathrm{d}z,\\
  &\left. \sigma \right|_{z=0 } =\frac{-\alpha \text{Re}\text{Ri} }{ r R } \left(  h (r)  + \sum\limits_{i=1}^{2}(\rho_{i}-1)\int\limits_{0}^{h (r)}\phi_{i}\mathrm{d}z \right),\\
  &\left. \sigma \right|_{z=h (r) }=0. \label{eq:equilibrium flux region 12e} 
\end{align}
\end{subequations}
where $\chi= \phi_{1}/(\phi_{1}+\phi_{2})$ is the ratio of the light particles to the total particles.  Based on the definition, $0 \leq \chi \leq 1$. Thus, given the fact $\sigma \leq 0$ and $\rho_{2} \geq \rho_{1}$, $\partial_{z}\chi$ is always nonnegative. This indicates that the ratio of the light particles continues to increase as $z$ increases. In other words, the lighter particles tend to concentrate near the free surface, while the heavier particles settle at the bottom of the channel. This holds true regardless of the ratio of the total volume fraction $\int_{0}^{h(r)}\phi_{1}\mathrm{d}z / \int_{0}^{h(r)}\phi_{2}\mathrm{d}z$. In addition, it is noteworthy that $\partial_{z}\chi$ is directly proportional to the density difference between two particle species. A larger density difference leads to a sharper transition from the region rich in heavy particles to the region rich in light particles. In cases where the density difference is minimal, the two particle species become thoroughly mixed, effectively reducing the scenario to that of a single-particle species. 
Following this argument, we consider the scenario where the secondary flow near the free surface transports heavier particles from region (\romannumeral 1) to region (\romannumeral 2). The heavier particles swiftly settle at the channel bottom, as depicted in panel (a) of Fig.~\ref{fig:criticalVolumeFractionBi}, and are then carried back to the higher-density particle-rich region (\romannumeral 1) by the secondary flow.
Furthermore, the secondary flow near the bottom of the channel transports lighter particles from region (\romannumeral 2) to region (\romannumeral 1). Panel (b) of Fig.~\ref{fig:criticalVolumeFractionBi} illustrates a scenario where a small amount of light particles is added to a region consisting of heavy particles at approximately the critical volume fraction. As the particles reach equilibrium in the vertical direction, lighter particles rise to the free surface. Subsequently, particles near the free surface are transported back to region (\romannumeral 2) by the secondary flow, restoring equilibrium. Therefore, the stability of the three-region stratification in the $r$-direction is justified.

With a similar analysis, one can show other configurations are not stable, for example, when the particles are concentrated in the outer region of the spiral as illustrated in panel (b) of Fig.~\ref{fig:stability}. Even if the system is initialized with an unstable configuration, it will converge to the stable configuration presented in panel (a) of Fig.~\ref{fig:stability}.

\subsection{Formula of flow and free surface profile}
In this section, we will derive the expression of fluid flow and free surface  in each region. Notice that in all three regions, the density and viscosity are functions of $r$ only.

From equation \eqref{eq:leading order equation d}, we integrate to solve for the pressure
\begin{equation}
  \begin{aligned}
   &p=  \frac{-\text{Re}\text{Ri}}{1+\frac{\alpha^{2}}{r^2 R^2} } \int\limits_{h (r)}^{z}\rho (r,s) \mathrm{d} s= \frac{\text{Re}\text{Ri} \rho (r) (h (r)-z)}{1+\frac{\alpha^{2}}{r^2 R^2} }.
\end{aligned}
\end{equation}
Now we can substitute pressure in equation \eqref{eq:leading order equation c} and with the associated boundary condition for $u_{\theta}$, we have 
\begin{align}
& \frac{\alpha \text{Re}\text{Ri} \rho}{ r^{2}R\left( 1+\frac{\alpha^{2}}{r^2 R^2} \right) }- \left(1+\frac{\alpha^{2}}{r^2 R^2}  \right)\partial_{z} (\mu \partial_{z}u_{\theta}) =0, \nonumber \\  
&\quad \left.  \partial_{z}u_{\theta} \right|_{z=h (r) }=0,\quad \left. u_{\theta} \right|_{z=0 }=0.
\end{align}
The angular velocity component is
\begin{equation}\label{eq:utheta}
\begin{aligned}
&u_{\theta}= \frac{\alpha }{r^2 R} \frac{ \text{Re} \text{Ri} \rho (r) z(z-2 h(r)) }{2 \mu (r)\left( 1+\frac{\alpha^{2}}{r^2 R^2} \right)^{2}}.\\
\end{aligned}
\end{equation}
We look for the radial velocity component by rearranging equation \eqref{eq:leading order equation b} to get the following equation and boundary conditions 
\begin{align}
\left(1+\frac{\alpha^{2}}{r^2 R^2}  \right)\partial_{z} (\mu \partial_{z}u_{r})&=\partial_{r}p -\frac{2\alpha \mu  }{r R}\partial_{z}u_{\theta}-r\text{Re}\rho   u_{\theta} ^2, \nonumber \\
\quad \left.  \partial_{z}u_{r} \right|_{z=h (r) }=0&,\; \left. u_{r} \right|_{z=0 }=0.
\end{align}
Substituting $u_{\theta}$ and solving yields the radial velocity component
\begin{align}\label{eq:ur expression1}
  u_{r}&=-\frac{\alpha ^2 \text{Re}^3 \text{Ri}^2 z \rho ^3 \left(z^5-2 h \left(-5 z^3 h+8 h^4+3 z^4\right)\right)}{120R^{2} \left( 1+ \frac{\alpha^2}{R^{2}r^{2}} \right) ^5 r^3 \mu ^3}\nonumber \\
  &-\frac{2 \alpha ^2 \text{Re} \text{Ri} z \rho  \left(-3 z h+3 h^2+z^2\right)}{3 R^{2} \left( 1+ \frac{\alpha^2}{R^{2}r^{2}} \right) ^3 r^3 \mu }\nonumber \\
  &-\frac{\text{Re} \text{Ri} z \left(\left(-3 z h+3 h^2+z^2\right) \partial_{r} \rho -3 \rho  (z-2 h)\partial_{r} h\right)}{6 \left( 1+ \frac{\alpha^2}{R^{2}r^{2}} \right)^2 \mu }.
\end{align}
Based on the continuity equation \eqref{eq:continuity nondimensionalization}, the vertical velocity component can be expressed as 
\begin{eqnarray}
u_{z}= -\int_0^{h (r)} \frac{1}{r}\partial_{r} (r u_{r})\mathrm{d} z.
\end{eqnarray}

The remaining task is to solve for the surface of the fluid in the equilibrium profile, $h (r)$. Substituting the expression of $u_{r}$ into $\int_0^{h (r)}u_{r}\mathrm{d}z=0$ yields
\begin{align}\label{eq:fluid depth}
\partial_{r}h =&-h(r) \left(\frac{3 \alpha ^2}{2 R^{2}\left( 1+ \frac{\alpha^2}{R^{2}r^{2}} \right)  r^3}+\frac{3 \partial_{r}\rho (r)}{8 \rho (r)}\right) \nonumber \\
&+\frac{6 \alpha ^2 \text{Re}^2 \text{Ri} h(r)^4 \rho (r)^2}{35 R^{2}\left( 1+ \frac{\alpha^2}{R^{2}r^{2}} \right) ^3 r^3 \mu (r)^2}.
\end{align}
We seek a solution with boundary condition $h (R_{i})=h_{i}$. Since it is a Bernoulli differential equation, we multiply $h (r)^{-4}$ on both side of equation  \eqref{eq:fluid depth}  and use the substitution $\xi (r)=h (r)^{-3}$, which transforms equation into the linear first-order differential equation
\begin{equation}
\begin{aligned}
&\partial_{r}\xi =\xi \left(\frac{9 \alpha ^2}{2 R^{2}\left( 1+ \frac{\alpha^2}{R^{2}r^{2}} \right)  r^3}+\frac{9 \partial_{r}\rho }{8 \rho }\right)-\frac{18 \alpha ^2 \text{Re}^2 \text{Ri}  \rho ^2}{35 R^{2}\left( 1+ \frac{\alpha^2}{R^{2}r^{2}} \right) ^3 r^3 \mu ^2}  ,\\
& \xi (R_{i})=h_{i}^{-3}.
\end{aligned}
\end{equation}
Integrating with respect to $r$ gives the solution
\begin{equation}\label{eq:fluid depth intermediate}
\begin{aligned}
  &\xi (r) = \left(\frac{\rho (r)}{\left(1+ \frac{\alpha^{2}}{R^{2}r^{2}}\right)^2}\right)^{\frac{9}{8}}\left( h_{i}^{-3}\left(\frac{\rho (R_{i})}{\left(1+ \frac{\alpha^{2}}{R^{2}R_{i}^{2}}\right)^2}\right)^{-\frac{9}{8}}- \right.\\
  &\left.\int\limits_{R_{i}}^{r} \left(\frac{\rho (s)}{\left(1+ \frac{\alpha^{2}}{R^{2}s^{2}}\right)^2}\right)^{-\frac{9}{8}} \frac{18 \alpha ^2 \text{Re}^2 \text{Ri}  \rho ^2}{35 R^{2}\left( 1+ \frac{\alpha^2}{R^{2}s^{2}} \right) ^3 s^3 \mu ^2}    \mathrm{d}s\right).
\end{aligned}
\end{equation}
In region  (\romannumeral 1), $\phi (r,z) =\phi_{2} (r,z) =\phi_{c,2} (r)$, $\rho (r) =1+ (\rho_{2}-1)\phi_{c,2} (r)$, $\mu (r)=\left( 1- \phi_{c,2} (r)/\phi_{m} \right)^{-2} $.   In region  (\romannumeral 2), $\phi (r,z) =\phi_{1} (r,z) =\phi_{c,1} (r)$, $\rho (r) =1+ (\rho_{1}-1)\phi_{c,1} (r)$, $\mu (r)=\left( 1- \phi_{c,1} (r)/\phi_{m} \right)^{-2} $. In those two regions, the exact formula of the integral in equation \eqref{eq:fluid depth intermediate} is unavailable so numerical methods are used. 

Region  (\romannumeral 3)  consists entirely of the clear fluid. $\rho=\mu=1$ due to the nondimensionalization. In this case, the exact expression of integral in equation \eqref{eq:fluid depth intermediate} is available and the fluid layer thickness is 
\begin{align}\label{eq:fluid depth fluid}
  h (r)=  &\left(1+ \frac{\alpha^{2}}{R^{2}r^{2}}\right)^{\frac{2}{3}}\left( h_{i}^{-3}\frac{\left(1+ \frac{\alpha^{2}}{R^{2}R_{i}^{2}}\right)^{\frac{9}{4}}}{\left(1+ \frac{\alpha^{2}}{R^{2}r^{2}}\right)^{\frac{1}{4}}}\right.\nonumber \\
  & \left.+\frac{36 \mathrm{Re}^{2}\mathrm{Ri}}{35} \left( 1- \frac{\left(1+ \frac{\alpha^{2}}{R^{2}R_{i}^{2}}\right)^{\frac{1}{4}}}{\left(1+ \frac{\alpha^{2}}{R^{2}r^{2}}\right)^{\frac{1}{4}}} \right)
  \right)^{-\frac{1}{3}}. 
\end{align}

We can eliminate $\partial_{r}h$ in equation \eqref{eq:ur expression1} using  equation \eqref{eq:fluid depth}
\begin{align}
  u_{r}= &-\frac{\text{Re} \text{Ri} z \left(-15 z h +6 h ^2+8 z^2\right) \left(\left( 1+ \frac{\alpha^2}{R^{2}r^{2}} \right)   r^3 \partial_{r}\rho  +4 \frac{\alpha^{2}}{R^{2}} \rho  \right)}{48 \left( 1+ \frac{\alpha^2}{R^{2}r^{2}} \right)  ^3 r^3 \mu  } \nonumber \\
  &-\frac{\alpha ^2 \text{Re}^3 \text{Ri}^2 z \rho  ^3 \left(-42 z^4 h +70 z^3 h ^2-72 z h ^4+32 h ^5+7 z^5\right)}{840 R^{2}\left( 1+ \frac{\alpha^2}{R^{2}r^{2}} \right) ^5 r^3 \mu  ^3}.
\end{align}

After obtaining the height profile of the free surface, we can confirm that the secondary flow is clockwise. The expansions of $u_r$ at $z=0$ is
\begin{align}
u_{r}= &\frac{-\text{Re} \text{Ri} z h ^2}{\left( 1+ \frac{\alpha^2}{R^{2}r^{2}} \right)  ^3 r^3 \mu  } \left( \frac{1}{8} \left(\left( 1+ \frac{\alpha^2}{R^{2}r^{2}} \right)   r^3 \partial_{r}\rho  + \frac{4\alpha^{2}\rho}{R^{2}}  \right) \right. \nonumber \\
& \left. +\frac{4 \alpha ^2 \text{Re}^2 \text{Ri} h ^3 \rho  ^3}{105  R^{2}\left( 1+ \frac{\alpha^2}{R^{2}r^{2}} \right) ^2 \mu  ^2} \right)+\mathcal{O} (z^{2}).
\end{align}
It is straightforward to demonstrate that $\partial_{r}\rho > 0$ holds for all $r > 0$ within each region, except at the locations of the interface $r = R_{p}, R_{f}$, where $\rho$ is discontinuous. Therefore $u_r<0$ near the bottom of the channel.  The expansion of $u_{r}$ at $z=h $ is
\begin{align}
u_{r} =&\frac{\text{Re} \text{Ri} h ^3}{24 \left( 1+ \frac{\alpha^2}{R^{2}r^{2}} \right)  ^3 r^3 \mu  } \left( \frac{\alpha ^2 \text{Re}^2 \text{Ri} h ^3 \rho  ^3}{7 R^{2}\left( 1+ \frac{\alpha^2}{R^{2}r^{2}} \right)^2 \mu  ^2} \right. \nonumber \\
&+ \left. \frac{1}{2}\left( 1+ \frac{\alpha^2}{R^{2}r^{2}} \right)   r^3 \partial_{r}\rho + \frac{2\alpha^{2}\rho}{R^{2}}   \right)+\mathcal{O} (z-h)
\end{align}
Therefore $u_{r}>0$ near the free surface. 

We introduce the stream function $\psi$, defined by
\begin{equation}
\begin{aligned}
&u_{r}=- \frac{1}{r}\partial_{z}\psi,\quad u_{z}=\frac{1}{r}\partial_{r}\psi.
\end{aligned}
\end{equation}
By requiring $\psi=0$ on the channel bottom $z=0$, we have
\begin{equation}\label{eq:stream function}
\begin{aligned}
\psi=&\frac{\text{Re} \text{Ri} z^2 (2 z-3 h ) (z-h ) \left(\left( 1+ \frac{\alpha^2}{R^{2}r^{2}} \right)   r^3 \partial_{r}\rho +\frac{4\alpha^{2}\rho}{R^{2}}  \right)}{48 \left( 1+ \frac{\alpha^2}{R^{2}r^{2}} \right)  ^3 r^2 \mu  }\\
&+\frac{\alpha ^2 \text{Re}^3 \text{Ri}^2 z^2 \rho  ^3 (z-2 h )^2 (z-h ) \left(z^2-2 h  (2 h +z)\right)}{840R^{2} \left( 1+ \frac{\alpha^2}{R^{2}r^{2}} \right)  ^5 r^2 \mu  ^3}. \\
\end{aligned}
\end{equation}
Note that the solution for the velocity components may not satisfy the no-slip boundary condition at the side wall. To address this, an additional boundary layer correction term is necessary to ensure compliance with the specified boundary conditions. \cite{stokes2004thin,arnold2017thin} Nevertheless, given the assumption that the channel width significantly exceeds the fluid depth, the boundary layer region occupies only a small fraction of the domain. In such cases, the thin-film approximation continues to be a reasonable and justifiable approach.

Given the initial height of the free surface at the inner wall of the channel, we can determine the free surface profile in region (\romannumeral 1). We make the assumption of a continuous free surface and utilize the solution's value at $r=R_{p}$ as the boundary condition to calculate the height profile in region (\romannumeral 2). Subsequently, employing the solution's value at $r=R_{f}$ as the boundary condition, we solve for the height profile in region (\romannumeral 3). The unknowns in this context are the height of the free surface at the inner wall of the channel, denoted as $h_{i}$, and the positions of the interface, namely $R_{p}$ and $R_{f}$. These parameters can be determined by evaluating the fluid flux and particle flux for each distinct particle species
\begin{equation}\label{eq:flow flux}
\begin{aligned}
&  Q_{f}=\int\limits_{R_{f}}^{R_{o}} \int\limits_{0}^{h (r)} r u_{\theta}\mathrm{d}z\mathrm{d}r=-\int\limits_{R_{f}}^{R_{o}}      \frac{\alpha }{r R} \frac{ \text{Re} \text{Ri} \rho (r) h^{3} (r) }{3 \mu (r)\left( 1+\frac{\alpha^{2}}{r^2 R^2} \right)^{2}} \mathrm{d}r,
  \\
  &Q_{p,1}=\int\limits_{R_{i}}^{R_{p}} \int\limits_{0}^{h (r)} r \phi_{1} u_{\theta}\mathrm{d}z\mathrm{d}r, \quad Q_{p,2}=\int\limits_{R_{p}}^{R_{f}} \int\limits_{0}^{h (r)} r \phi_{2} u_{\theta}\mathrm{d}z\mathrm{d}r.
\end{aligned}
\end{equation}
Given a specific flux, we can solve for $h_{i}$, $R_{p}$ and $R_{f}$ using numerical method. If one of the particle fluxes is zero, then the current model reduces to the single-particle species model. \cite{lee2014behavior, arnold2019particle}

\subsection{Scaling}
Here we analyze the magnitude of each quantity in the formulas from the previous section. 
 This analysis provides insight into the selection of a characteristic fluid layer thickness and characteristic velocity.
Following the ideas in\cite{stokes2013thin,arnold2015thin}, to ensure the coefficient of $h^{4}$ in equation \eqref{eq:fluid depth fluid} has unit value at the midpoint of the channel $r=R_{i}'+\frac{1}{2}$, we estimate the dimensionless fluid depth as
\begin{equation}
\begin{aligned}
&h'\sim  \left( \frac{ 6\alpha ^2 \text{Re}^2 \text{Ri}}{35R^2} \right)^{-\frac{1}{3}} \left(R_{i}'+\frac{1}{2}\right) \left(\frac{ \alpha ^2}{R^{2}\left( R_{i}'+\frac{1}{2}\right){}^2}+1\right) .
\end{aligned}
\end{equation}
With the change of variable for the nondimensionalization \eqref{eq:nondimensionalization}, the scaling of the dimensional fluid depth is
\begin{equation}\label{eq:characteristic fluid depth}
\begin{aligned}
&h=Hh'\sim \sqrt[3]{\frac{\mu_{\ell} ^2}{g \rho_{\ell} ^2}} \sqrt[3]{\frac{35}{6\alpha^{2}R}}  \left(R_{i}+\frac{R}{2}\right) \left(\frac{ \alpha ^2}{\left( R_{i}+\frac{R}{2}\right){}^2}+1\right) .
\end{aligned}
\end{equation}

The value of equation \eqref{eq:utheta} at the midpoint of the channel implies the scaling of the dimensionless angular velocity is
\begin{equation}
\begin{aligned}
&u_{\theta}'=  \frac{\alpha \text{Re} \text{Ri} h' (r)^{2}}{2 R } \frac{\left(R_{i}'+\frac{1}{2}\right){}^2}{\left(\left(R_{i}'+\frac{1}{2}\right){}^2+\frac{\alpha ^2}{R^2}\right){}^2}.\\
\end{aligned}
\end{equation}
The scaling of the dimensional angular velocity is
\begin{equation}\label{eq:characteristic angular velocity}
\begin{aligned} 
&u_{\theta}=\frac{U}{R}u_{\theta}'\sim \frac{1}{2}\left(\frac{35}{6}\right)^{2/3} \sqrt[3]{\frac{g \mu_{\ell} }{  \rho_{\ell}  \alpha R^2}}.
\end{aligned}
\end{equation}
We can take its reciprocal as the time scale
\begin{equation}
  \begin{aligned}
    T\sim 2\left(\frac{6}{35}\right)^{2/3} \sqrt[3]{\frac{ \rho_{\ell}  \alpha R^2  }{g \mu_{\ell} }}.
\end{aligned}
\end{equation}

Equation \eqref {eq:characteristic fluid depth} and \eqref{eq:characteristic angular velocity} appear to suggest that a larger viscosity leads to a thicker fluid layer and a larger velocity. This is counterintuitive, but can be justified by noticing that the flow flux increases as viscosity increases in this setup. We can show from equation \eqref{eq:flow flux} that the characteristic scale of the dimensionless flux is
\begin{equation}
\begin{aligned}
&Q'\sim \frac{35R}{18\alpha  \text{Re}} \left(R_{i}'+\frac{1}{2}\right)^{3} \left(\frac{ \alpha ^2}{R^{2}\left( R_{i}'+\frac{1}{2}\right){}^2}+1\right). \\
\end{aligned}
\end{equation}
The scaling of the dimensional flow flux is 
\begin{equation}
\begin{aligned}
&Q=UHRQ'\sim \frac{35 \mu_{\ell} }{18\alpha R  \rho_{\ell}} \left(R_{i}+\frac{R}{2}\right)^{3} \left(\frac{ \alpha ^2}{\left( R_{i}+\frac{R}{2}\right)^2}+1\right).\\
\end{aligned}
\end{equation}
Thus, when the characteristic scale of the flow flux is fixed, a larger viscosity leads to a smaller velocity and a thinner fluid layer. This is consistent with the expected physics.

\section{Complete steady state solution}\label{sec:steadystate}

\begin{figure*}
  \centering
  \subfigure[ ~Apparatus 1 profile]{
    \includegraphics[width=0.47\linewidth]{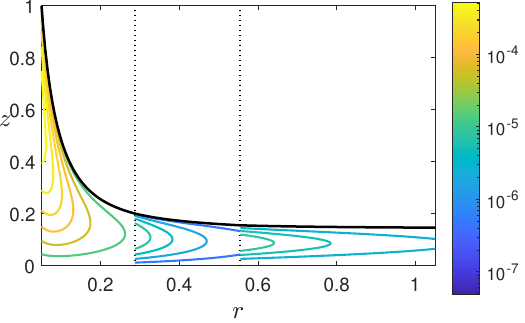}
  }
\subfigure[ ~Apparatus 2 profile]{
    \includegraphics[width=0.47\linewidth]{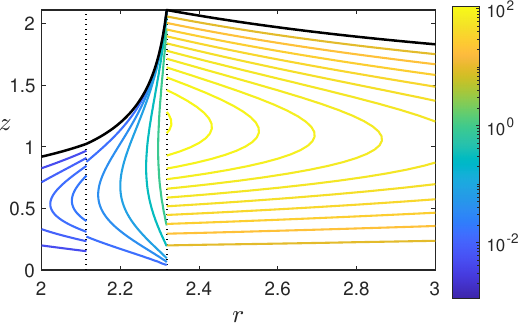}
  }
    \subfigure[ ~Apparatus 1 particle volume fraction]{
    \includegraphics[width=0.47\linewidth]{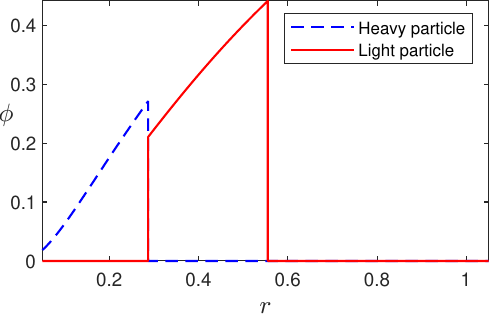}
  }
\subfigure[ ~Apparatus 2 particle volume fraction]{
    \includegraphics[width=0.47\linewidth]{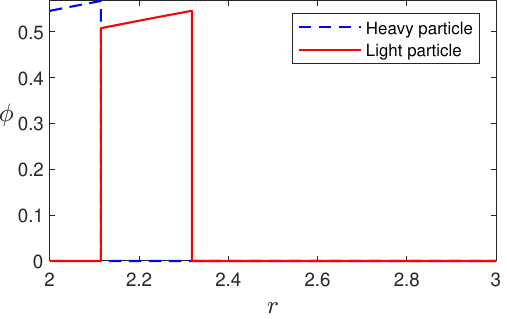}
  }
   \subfigure[ ~Apparatus 1 geometry]{
    \includegraphics[width=0.47\linewidth]{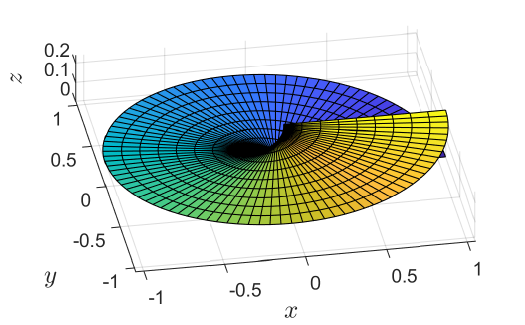}
  }
   \subfigure[ ~Apparatus 2 geometry]{
    \includegraphics[width=0.47\linewidth]{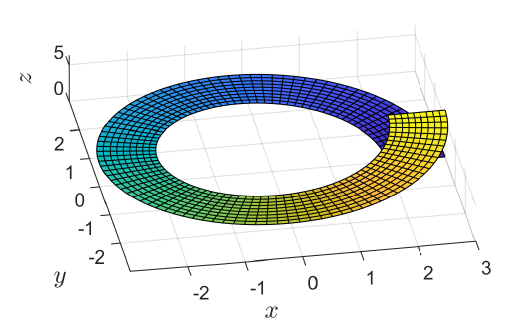}
  }
  \hfill
  \caption[]
  {Equilibrium profiles and geometries for two different spiral concentrator setups (see parameters in Table \ref{table: parameter 1}).  Row (a,b) show the fluid depth $h$ as functions of $r$ and the stream function $\psi$ defined in equation \eqref{eq:stream function}. The vertical black dotted lines indicate the location of the interface of each region (i.e. left is $R_p$ and right is $R_f$). Row (c,d) show the particle volume fractions $\phi$ as functions of $r$. The red solid line represents the light particles $\phi_1$ and the blue dash line represents the heavy particles $\phi_2$. The particle volume fraction $\phi_{c,i}$ is computed by equation \eqref{eq:critical volume fraction}.  The plots (e) and (f) show their respective geometries with respect to dimensionless variables.}
  \label{fig:freeSurface}
\end{figure*}

\begin{table} 
\caption{\label{table: parameter 1} Practical physical parameters for a spiral separator.} 
\begin{ruledtabular}
\begin{tabular}{lll}
\textbf{Variable} & \textbf{Apparatus 1\cite{jxscmachine2024table}} & \textbf{Apparatus 2} \\
\hline
  R & $0.2 \, \text{m}$ & $0.5 \, \text{m}$ \\
 $\alpha$ & 0.0382 \text{m} &1 \text{m} \\ 
$R_{i}$ & $0.05 \, \text{m}$ & $2 \, \text{m}$ \\
$H$ & $4.48 \times 10^{-4} \, \text{m}$ & $5.2 \times 10^{-4} \, \text{m}$ \\
$U$ & $0.228 \, \text{m/s}$ & $0.45 \, \text{m/s}$ \\
$\mu$ & $10^{-3} \, \text{kg} \cdot \text{m}^{-1} \cdot \text{s}^{-1}$ & - \\
$\rho_{\ell}$ & $10^{3} \, \text{kg} \cdot \text{m}^{-3}$ & - \\
$g$ & $9.81 \, \text{m} \cdot \text{s}^{-2}$ & - \\
$\rho_{2}$ & $3.8 \rho_{\ell}$ & - \\
$\rho_{1}$ & $2.5 \rho_{\ell}$ & - \\
  $d$ & $5 \times 10^{-5} \, \text{m}$ & - \\
\end{tabular}
\end{ruledtabular}
\end{table}

We consider two flow configurations with the practical physical parameters in Table \ref{table: parameter 1}. We first analyze the parameters in Apparatus 1 of Table \ref{table: parameter 1}, where the inner radius is notably small relative to the channel width, as shown in Fig.~\ref{fig:freeSurface}(e).  This configuration reflects the design of real spiral separators, which are constructed compactly with small inner radii.\cite{jxscmachine2024table}  We have $\epsilon= \frac{H}{R}\approx 0.001$  and $\frac{d}{H}\approx 0.12$, which satisfies the assumption of the thin-film approximation.  The resulting dimensionless parameters are $\mathrm{Re}=102.094$,  $\mathrm{Ri}=0.0845$. With given $Q_{p1}/Q_{f}=0.3$ and $Q_{p2}/Q_{f}=2$, we have $h_{i}=1$, $R_{p}=R_{i}+0.237$ and $R_{f}=R_{i}+0.505$. The solution of the fluid depth and volume fraction is presented in Fig.~\ref{fig:freeSurface}(a). In the plot, the fluid depth is largest at the inner side. The particle volume fraction is not continuous at the location of the interface between each region, as we discussed before. Since the flux of the heavier particle is larger then the flux of the light particle, the value of $\phi_2$ in  Region (\romannumeral 1) is lower  than $\phi_1$ in region (\romannumeral 2) and region (\romannumeral 2) is wider than region (\romannumeral 1). In both regions, the concentration is highest towards the outer side of the region.

Next we focus on Apparatus 2 where the inner radius is larger compared with the channel width (see Fig.~\ref{fig:freeSurface}(f)). The parameters are in the third column of Table \ref{table: parameter 1}.
The resulting dimensionless parameters are $\mathrm{Re}=233.3$,  $\mathrm{Ri}=0.025$. With given $Q_{p1}/Q_{f}=0.005$ and $Q_{p2}/Q_{f}=0.001$, we have $h_{i}=0.9196$, $R_{p}=R_{i}+0.114$ and $R_{f}=R_{i}+0.318$. The solution of the fluid depth and stream functions is presented in Fig.~\ref{fig:freeSurface}(b). Interestingly, in this scenario, the fluid depth does not follow a monotonic pattern. The maximum fluid depth occurs at $R_f$ the interface between the lighter particle and the clear fluid. The particle volume fractions in each region are shown in ~\ref{fig:freeSurface}(d). Region (\romannumeral 1) contains $\phi_2$ at a higher concentration than $\phi_1$ in region (\romannumeral 2). Again, region (\romannumeral 2) is wider than region (\romannumeral 1).

\section{Conclusion}
\label{sec:conclusion}
\begin{figure}
    \centering
    \includegraphics[width=\linewidth]{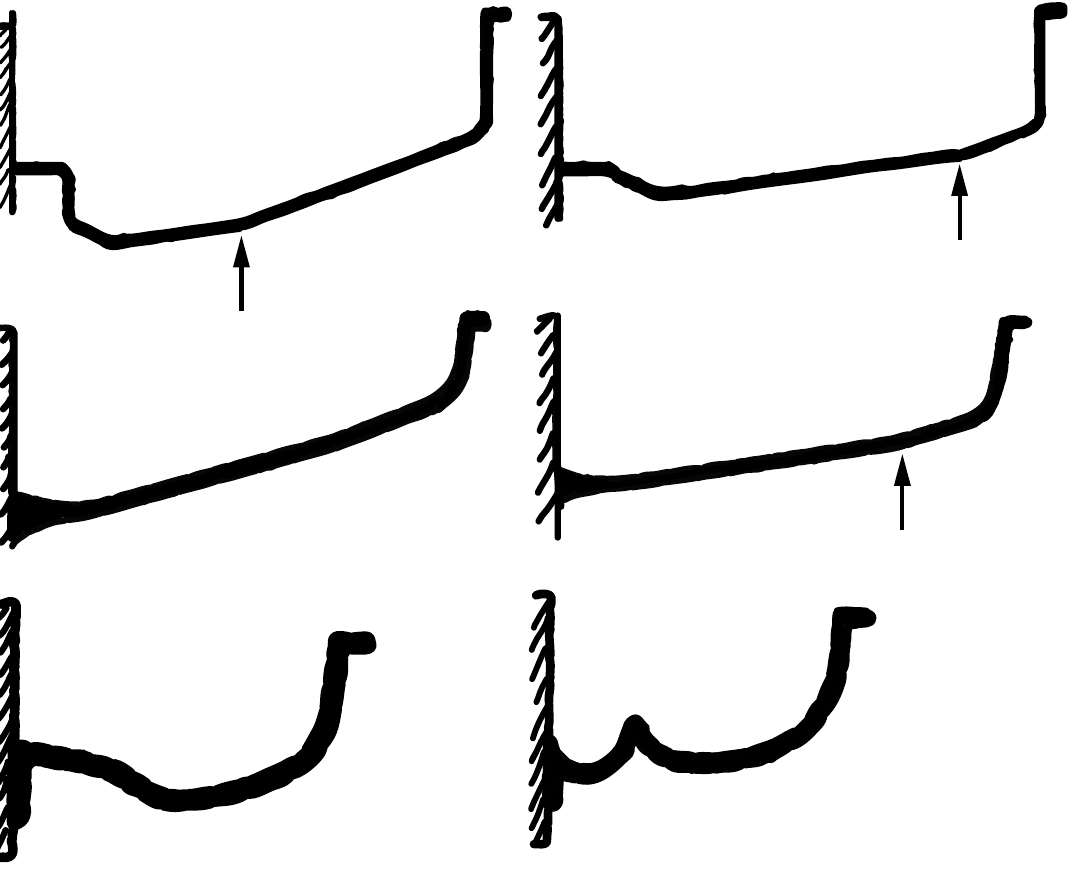}
    \caption{Sketches of the cross sections on spiral separators from various patents. The first column are diagrams taken from cross sections near the middle of the separator and the second column are cross sections near the exit of the device. The first row are from Giffard 1982\cite{giffardpatent}, the second row are from Wright 1986\cite{WrightpatentUS}, and the third row are from Wright 1985\cite{WrightpatentEU}. The arrows highlight the location where the slopes of the bottom change. }
    \label{fig:patent_sketches}
\end{figure}
In this study, we successfully apply our model to investigate the behavior of gravity-driven bidensity slurries on an inclined spiral separator, taking into account particle interactions through empirically derived formulas. Specifically, we developed a reduced model for velocity and particle volume fraction, specifically applicable when the fluid depth is significantly smaller than the channel width of the spiral separator.
Our comprehensive analysis uncovered a distinctive radial particle stratification within the spiral separator. As illustrated in Fig.~\ref{fig:stability}, the entire cross-section of the spiral separator can be discretized into three non-overlapping regions. The innermost region, adjacent to the inner side wall, is predominantly comprised of heavy particles. The middle region is a mixture of fluid with light particles, while the outermost region, in close proximity to the outer sidewall, is primarily particle-free fluid.
Having identified this radial stratification, we derive expressions for the velocity field, particle distribution, and fluid depth. Additionally, we demonstrate solutions for physical parameters used in practice, enhancing the applicability and utility of our findings in real-world scenarios. Adjustable splitters can be placed at the exit of the apparatus so that the different slurry components split into individual streams.\cite{boisvert2023axial,giffardpatent} Knowing the $R_p$ and $R_f$, can inform positioning of adjustable splitters prior to feeding material into the apparatus. Improved sorting can reduce the number of stages where the output slurry must be reprocessed.\cite{boisvert2023axial} This research paves a way for enhanced efficiency and optimization in related industrial processes.

Future research directions encompass several avenues. First, in this study, the thickness of the interface between different regions is assumed to be zero. In reality, there exists a non-zero transition layer, which is especially pronounced when the density difference is small. It is intriguing to explore the particle distribution within this transition layer. Second, this study focuses on the spiral separator with a flat bottom. Patents show the channel bottom tilts at an angle with the outer wall higher than the inner wall as can be seen in the sketches in Fig. \ref{fig:patent_sketches}. In practice, the spiral concentrators have cross sections which change in topology from the top to bottom. This is in effort to keep the mixture initially well-mixed and then enhance separation of the slurry components as the mixture moves towards the exit. In the figure, we look at cross sections towards the middle and bottom where ideally we have achieved equilibrium. The left column are cross sections from the middle of the channel and the right column are cross sections near the exit. 
Therefore an improvement would be to generalize the results to channels with arbitrary topography. For certain topographies, the secondary flow could be anticlockwise, leading to a different stable configuration. Third, this study concentrates on the steady flow solution with equilibrium particle distribution. We would like to investigate the transient and how the system converges to this steady state. Furthermore, we would like to understand how to the change in cross-section topology from the middle of the separator to the exit influences the final equilibrium solution. Our work establishes the fundamental knowledge required to further optimize the design of the spiral concentrator to accelerate separation.

\section{Acknowledgments}
This material is based upon work supported by the U.S. National Science Foundation under award No. DMS-2407006. This work is also supported by Simons Math + X Investigator Award number 510776. Sarah C. Burnett was supported by the 2022 L'Or\'eal USA for Women in Science Postdoctoral Fellowship.

\appendix

\section{Helical coordinates system}\label{sec: app Helical coordinates system}

Notice that $\mathbf{e}_{r}$ is orthogonal to $\mathbf{e}_{\theta}$ and $\mathbf{e}_{z}$, but $\mathbf{e}_{z}$ is not orthogonal to $\mathbf{e}_{\theta}$. The formula for the  dot product becomes
\begin{align} \label{eq:non_ortho_dot}
(f_{r}&\mathbf{e}_{r}+f_{\theta}\mathbf{e}_{\theta}+f_{z}\mathbf{e}_{z})\cdot(g_{r}\mathbf{e}_{r}+g_{\theta}\mathbf{e}_{\theta}+g_{z}\mathbf{e}_{z}) \nonumber \\
&= f_{r}g_{r}+(r^{2}+\alpha^{2})f_{\theta}g_{\theta}+f_{z}g_{z}+ \alpha (f_{\theta}g_{z}+f_{z}g_{\theta}).
\end{align}

We can apply the chain rule to a vector $\mathbf{f}$ and use the position from equation \eqref{eq: position} to obtain the following conversions between the differential operators:
\begin{equation} \label{eq: partial_rtz}
\begin{cases}
\begin{aligned}
  \partial_{r}&=\cos \theta \partial_{x}+ \sin \theta \partial_{y}, \;\\
  \partial_{\theta}&=-r \sin \theta \partial_{x}+r\cos \theta \partial_{y}+\alpha \partial_{z}, \; \\
  \partial_{z}&=\partial_{z},\\
\end{aligned}
\end{cases}
\end{equation}

\begin{equation} \label{eq: partial_xyz}
\begin{cases}
\begin{aligned}
  \partial_{x}&=\cos \theta \partial_{r}-\frac{\sin \theta}{r} \partial_{\theta}+\frac{\alpha \sin \theta}{r}\partial_{z}, \; \\
  \partial_{y}&=\sin \theta\partial_{r}+\frac{\cos \theta}{r} \partial_{\theta}-\frac{\alpha \cos \theta}{r}\partial_{z} , \; \\
  \partial_{z}&=\partial_{z}. 
\end{aligned}
\end{cases}
\end{equation} 
The conversion between the gradient in two different coordinate systems is found by substituting equation \eqref{eq: basis vectors cart} into $\nabla f= \partial_{x}f\mathbf{e}_{x}+\partial_{y}f \mathbf{e}_{y}+\partial_{z}f \mathbf{e}_{z}$ and collecting the coefficients. Then the formula for the gradient is
\begin{equation} \label{eq: gradient}
\nabla f=\partial_{r}f \mathbf{e}_{r}+\frac{1}{r^{2}}  \left( \partial_{\theta}f-\alpha\partial_{z} f\right)\mathbf{e}_{\theta}+\left( \frac{-\alpha}{r^{2}}\partial_{\theta}f+ \left( 1+ \frac{\alpha^{2}}{r^{2}} \right)\partial_{z}f \right)\mathbf{e}_{z}.
\end{equation} 

To compute the shear rate, we need $\nabla f\cdot \nabla f$. Take the dot product of equation \eqref{eq: gradient} with itself using the relation in equation \eqref{eq:non_ortho_dot}. Then the formula for $\nabla f\cdot \nabla f$ is
\begin{equation}
\nabla f\cdot \nabla f=\left( \partial_{z}f \right)^{2}+ \left( \frac{\alpha \partial_{z}f}{r} \right)^{2}- \frac{2\alpha \partial_{z}f  \partial_{\theta}f}{r^{2}}+ \left( \frac{\partial_{\theta}f}{r} \right)^{2}+ \left( \partial_{r}f \right)^{2}
\end{equation}
The conversion between the divergence in two different coordinate systems can be calculated by substituting in equations \eqref{eq: partial_xyz} and \eqref{eq: u_xyz} into $\nabla\cdot \mathbf{f}=\partial_{x}f_{x}+\partial_{y}f_{y}+\partial_{z}f_{z}$ and simplifying to get the form
\begin{equation} \label{eq: divergence}
\nabla\cdot \mathbf{f}=\frac{1}{r}\partial_{r}\left( rf_{r} \right)+ \partial_{\theta} f_{\theta}+\partial_zf_{z},
\end{equation}
where $\mathbf{f}=f_{x}\mathbf{e}_{x}+f_{y}\mathbf{e}_{y}+f_{z}\mathbf{e}_{z}=f_{r}\mathbf{e}_{r}+f_{\theta}\mathbf{e}_{\theta}+f_{z}\mathbf{e}_{z}$. To compute the Laplacian operator apply the divergence \eqref{eq: divergence} to the gradient \eqref{eq: gradient} to get 
\begin{equation} \label{eq: laplacian}
\Delta f=\frac{1}{r}\partial_{r}\left( r \partial_{r}f \right)+ \frac{1}{r^{2}}\partial_{\theta}^{2}f- \frac{2\alpha }{r^{2}}\partial_{\theta}\partial_{z}f+ \frac{\alpha^{2}+r^{2}}{r^{2}}\partial_{z}^{2}f .
\end{equation}
Next, the advection operator can be calculated by taking the dot product of a helical velocity vector and equation \eqref{eq: gradient} using the relation \eqref{eq:non_ortho_dot}. The result is
\begin{equation}
\begin{aligned}
\mathbf{u}\cdot\nabla f
&=\partial_{r}f u_{r}+  \left( \partial_{\theta}f-\alpha\partial_{z} f\right) \frac{(r^{2}+\alpha^{2}) u_{\theta}+ \alpha u_{z}}{r^{2}} \\
&\qquad \qquad + \frac{ (\alpha^{2}+r^{2})\partial_{z} f-\alpha\partial_{\theta}f}{r^{2}}\left( u_{z}+\alpha u_{\theta}\right). 
\end{aligned}
\end{equation}

\section{The asymptotic expansion}\label{sec: app expansion}

The dimensionless Navier-Stokes equation before taking $\epsilon\rightarrow 0$ is 
\begin{widetext}
\begin{equation}
\begin{aligned}\label{eq: full NS}
\mathrm{Re}&\rho \left(\epsilon^{2} u_{r}\partial_{r}u_{r}+\epsilon^{2} u_{z} \partial_{z}u_{r}-  ru_{\theta} u_{\theta} \right)=- \partial_{r}p+ 2\epsilon^{2}\partial_{r}\mu \partial_{r}u_{r}  +\mu \left( \frac{\epsilon^{2}}{r}\partial_{r}\left( r \partial_{r}u_{r} \right)+ \left( 1+ \frac{\alpha^{2}}{R^{2}} \frac{1}{r^{2}} \right)\partial_{z}^{2}u_{r} -  \frac{\epsilon^{2}u_{r}}{r^{2}} + \frac{\alpha}{R}\frac{2}{r} \partial_{z} u_{\theta} \right),  \\
 &\hspace{1cm}+\partial_{z}\mu \left( \left( 1+ \frac{\alpha^{2}}{R^{2}}\frac{1}{r^{2}} \right) \partial_{z} u_{r}+\epsilon^{2}\partial_{r}u_{z}\right) \\
\epsilon^{2} &\mathrm{Re}\rho \left(  u_{r} \partial_{r}u_{\theta}+  u_{z} \partial_{z}u_{\theta}+ \frac{2}{r} u_{r} u_{\theta} \right)= \partial_{z}\mu \left( - \frac{\epsilon^{2}\alpha}{R}\frac{2 u_{r}}{r^{3}} +\left( 1+\frac{\alpha^{2}}{R^{2}}\frac{1}{r^{2}} \right) \partial_{z} u_{\theta}- \frac{\epsilon^{2}\alpha}{R} \frac{1 }{r^{2}}\partial_{z} u_{z} \right)  
 \\
&\qquad + \frac{\alpha}{R}\frac{1}{r^{2}} \partial_{z}p+\mu  \left( \epsilon^{2} \partial_{r}^{2} u_{\theta}+  \left( 1+ \frac{\alpha^{2}}{R^{2}} \frac{1}{r^{2}} \right) \partial_{z}^{2} u_{\theta} +\epsilon^{2}\frac{3}{r} \partial_{r} u_{\theta}-  \frac{\alpha \epsilon^{2}}{R}\frac{2  }{r^{3}}\partial_{z} u_{r}  \right) +\epsilon^{2} \partial_{r}\mu \left( -  \frac{\alpha}{R}\frac{1 }{r^{2}} \partial_{z} u_{r}+\partial_{r}u_{\theta} \right)  \\
\epsilon^{2} &\mathrm{Re}\rho \left(\epsilon^{2} u_{r}\partial_{r}u_{z}+ \epsilon^{2}u_{z}\partial_{z}u_{z}- \frac{\alpha}{R}\frac{2}{r} u_{r}u_{\theta} \right) = - \left( 1+ \frac{\alpha^{2}}{R^{2}}\frac{1}{r^{2}} \right) \partial_{z}p+\epsilon^{2}\mu \left(\epsilon^{2}\partial_{r}^{2}u_{z}+ \left( 1+ \frac{\alpha^{2}}{R^{2}}\frac{1}{r^{2}} \right) \partial_{z}^{2}u_{z}+ \frac{ \alpha^{2}}{R^{2}}\frac{2 }{r^{3}} \partial_{z}u_{r} + \epsilon^{2}\frac{1}{r}\partial_{r}u_{z} \right.  \\
 & \left.- \frac{\alpha}{R} \frac{2}{r} \partial_{r}u_{\theta} \right)+ \epsilon^{2}\partial_{r}\mu  \left( \left( 1+ \frac{\alpha^{2}}{R^{2}}\frac{1}{r^{2}} \right) \partial_{z}u_{r}+\epsilon^{2}\partial_{r}u_{z} \right)  +2 \epsilon^{2} \partial_{z}\mu \left( \frac{ \alpha^{2}}{R^{2}}\frac{1}{r^{3}}u_{r}+ \left( 1+ \frac{\alpha^{2}}{R^{2}}\frac{1}{r^{2}} \right)\partial_{z}u_{z} \right)- \mathrm{Re}  \mathrm{Ri}\rho.
\end{aligned}
\end{equation}
\end{widetext}

\nocite{*}
%

\begin{thebibliography}{45}%
\makeatletter
\providecommand \@ifxundefined [1]{%
 \@ifx{#1\undefined}
}%
\providecommand \@ifnum [1]{%
 \ifnum #1\expandafter \@firstoftwo
 \else \expandafter \@secondoftwo
 \fi
}%
\providecommand \@ifx [1]{%
 \ifx #1\expandafter \@firstoftwo
 \else \expandafter \@secondoftwo
 \fi
}%
\providecommand \natexlab [1]{#1}%
\providecommand \enquote  [1]{``#1''}%
\providecommand \bibnamefont  [1]{#1}%
\providecommand \bibfnamefont [1]{#1}%
\providecommand \citenamefont [1]{#1}%
\providecommand \href@noop [0]{\@secondoftwo}%
\providecommand \href [0]{\begingroup \@sanitize@url \@href}%
\providecommand \@href[1]{\@@startlink{#1}\@@href}%
\providecommand \@@href[1]{\endgroup#1\@@endlink}%
\providecommand \@sanitize@url [0]{\catcode `\\12\catcode `\$12\catcode
  `\&12\catcode `\#12\catcode `\^12\catcode `\_12\catcode `\%12\relax}%
\providecommand \@@startlink[1]{}%
\providecommand \@@endlink[0]{}%
\providecommand \url  [0]{\begingroup\@sanitize@url \@url }%
\providecommand \@url [1]{\endgroup\@href {#1}{\urlprefix }}%
\providecommand \urlprefix  [0]{URL }%
\providecommand \Eprint [0]{\href }%
\providecommand \doibase [0]{https://doi.org/}%
\providecommand \selectlanguage [0]{\@gobble}%
\providecommand \bibinfo  [0]{\@secondoftwo}%
\providecommand \bibfield  [0]{\@secondoftwo}%
\providecommand \translation [1]{[#1]}%
\providecommand \BibitemOpen [0]{}%
\providecommand \bibitemStop [0]{}%
\providecommand \bibitemNoStop [0]{.\EOS\space}%
\providecommand \EOS [0]{\spacefactor3000\relax}%
\providecommand \BibitemShut  [1]{\csname bibitem#1\endcsname}%
\let\auto@bib@innerbib\@empty
\bibitem [{\citenamefont {Lee}, \citenamefont {Stokes},\ and\ \citenamefont
  {Bertozzi}(2014)}]{lee2014behavior}%
  \BibitemOpen
  \bibfield  {author} {\bibinfo {author} {\bibfnamefont {S.}~\bibnamefont
  {Lee}}, \bibinfo {author} {\bibfnamefont {Y.}~\bibnamefont {Stokes}},\ and\
  \bibinfo {author} {\bibfnamefont {A.~L.}\ \bibnamefont {Bertozzi}},\
  }\bibfield  {title} {\enquote {\bibinfo {title} {Behavior of a particle-laden
  flow in a spiral channel},}\ }\href@noop {} {\bibfield  {journal} {\bibinfo
  {journal} {Physics of Fluids}\ }\textbf {\bibinfo {volume} {26}} (\bibinfo
  {year} {2014})}\BibitemShut {NoStop}%
\bibitem [{\citenamefont {Arnold}, \citenamefont {Stokes},\ and\ \citenamefont
  {Green}(2019)}]{arnold2019particle}%
  \BibitemOpen
  \bibfield  {author} {\bibinfo {author} {\bibfnamefont {D.}~\bibnamefont
  {Arnold}}, \bibinfo {author} {\bibfnamefont {Y.}~\bibnamefont {Stokes}},\
  and\ \bibinfo {author} {\bibfnamefont {J.}~\bibnamefont {Green}},\ }\bibfield
   {title} {\enquote {\bibinfo {title} {Particle-laden thin-film flow in
  helical channels with arbitrary shallow cross-sectional shape},}\ }\href@noop
  {} {\bibfield  {journal} {\bibinfo  {journal} {Physics of Fluids}\ }\textbf
  {\bibinfo {volume} {31}} (\bibinfo {year} {2019})}\BibitemShut {NoStop}%
\bibitem [{\citenamefont {Wong}\ and\ \citenamefont
  {Bertozzi}(2016)}]{wong2016conservation}%
  \BibitemOpen
  \bibfield  {author} {\bibinfo {author} {\bibfnamefont {J.~T.}\ \bibnamefont
  {Wong}}\ and\ \bibinfo {author} {\bibfnamefont {A.~L.}\ \bibnamefont
  {Bertozzi}},\ }\bibfield  {title} {\enquote {\bibinfo {title} {A conservation
  law model for bidensity suspensions on an incline},}\ }\href@noop {}
  {\bibfield  {journal} {\bibinfo  {journal} {Physica D: Nonlinear Phenomena}\
  }\textbf {\bibinfo {volume} {330}},\ \bibinfo {pages} {47--57} (\bibinfo
  {year} {2016})}\BibitemShut {NoStop}%
\bibitem [{\citenamefont {Kerns}\ and\ \citenamefont
  {Egbert}(1932)}]{Kernpatent}%
  \BibitemOpen
  \bibfield  {author} {\bibinfo {author} {\bibfnamefont {F.~W.}\ \bibnamefont
  {Kerns}}\ and\ \bibinfo {author} {\bibfnamefont {W.}~\bibnamefont {Egbert}},\
  }\bibfield  {title} {\enquote {\bibinfo {title} {Method and means for
  removing sand and the like from fluids ({US Patent} 1880185)},}\ }\href@noop
  {} {\bibfield  {journal} {\bibinfo  {journal} {United State Patent Office}\ }
  (\bibinfo {year} {1932})}\BibitemShut {NoStop}%
\bibitem [{\citenamefont {Wright}(1981)}]{WrightpatentEU}%
  \BibitemOpen
  \bibfield  {author} {\bibinfo {author} {\bibfnamefont {D.~C.}\ \bibnamefont
  {Wright}},\ }\href@noop {} {\enquote {\bibinfo {title} {A spiral separator
  ({EU Patent} 0039139).}}\ } (\bibinfo {year} {1981})\BibitemShut {NoStop}%
\bibitem [{\citenamefont {Giffard}(1982)}]{giffardpatent}%
  \BibitemOpen
  \bibfield  {author} {\bibinfo {author} {\bibfnamefont {P.~J.}\ \bibnamefont
  {Giffard}},\ }\bibfield  {title} {\enquote {\bibinfo {title} {Spiral
  separator ({WO Patent} 1982003187)},}\ }\href@noop {} {\bibfield  {journal}
  {\bibinfo  {journal} {World Intellectual Property Organization}\ } (\bibinfo
  {year} {1982})}\BibitemShut {NoStop}%
\bibitem [{\citenamefont {Wright}(1986)}]{WrightpatentUS}%
  \BibitemOpen
  \bibfield  {author} {\bibinfo {author} {\bibfnamefont {D.~C.}\ \bibnamefont
  {Wright}},\ }\bibfield  {title} {\enquote {\bibinfo {title} {Spiral
  separators ({US Patent} 4563279)},}\ }\href@noop {} {\bibfield  {journal}
  {\bibinfo  {journal} {United State Patent Office}\ } (\bibinfo {year}
  {1986})}\BibitemShut {NoStop}%
\bibitem [{\citenamefont {Weldon}(1997)}]{weldon1997fine}%
  \BibitemOpen
  \bibfield  {author} {\bibinfo {author} {\bibfnamefont {B.}~\bibnamefont
  {Weldon}},\ }\bibfield  {title} {\enquote {\bibinfo {title} {Fine coal
  beneficiation: Spiral separators in the australian industry},}\ }\href@noop
  {} {\bibfield  {journal} {\bibinfo  {journal} {Australian Coal Review}\ ,\
  \bibinfo {pages} {25--28}} (\bibinfo {year} {1997})}\BibitemShut {NoStop}%
\bibitem [{\citenamefont {Holland-Batt}\ and\ \citenamefont
  {Holtham}(1991)}]{holland1991particle}%
  \BibitemOpen
  \bibfield  {author} {\bibinfo {author} {\bibfnamefont {A.}~\bibnamefont
  {Holland-Batt}}\ and\ \bibinfo {author} {\bibfnamefont {P.}~\bibnamefont
  {Holtham}},\ }\bibfield  {title} {\enquote {\bibinfo {title} {Particle and
  fluid motion on spiral separators},}\ }\href@noop {} {\bibfield  {journal}
  {\bibinfo  {journal} {Minerals engineering}\ }\textbf {\bibinfo {volume}
  {4}},\ \bibinfo {pages} {457--482} (\bibinfo {year} {1991})}\BibitemShut
  {NoStop}%
\bibitem [{\citenamefont {Sayaslan}(2004)}]{sayaslan2004wet}%
  \BibitemOpen
  \bibfield  {author} {\bibinfo {author} {\bibfnamefont {A.}~\bibnamefont
  {Sayaslan}},\ }\bibfield  {title} {\enquote {\bibinfo {title} {Wet-milling of
  wheat flour: industrial processes and small-scale test methods},}\
  }\href@noop {} {\bibfield  {journal} {\bibinfo  {journal} {LWT-Food Science
  and Technology}\ }\textbf {\bibinfo {volume} {37}},\ \bibinfo {pages}
  {499--515} (\bibinfo {year} {2004})}\BibitemShut {NoStop}%
\bibitem [{\citenamefont {Majekodunmi}(2015)}]{majekodunmi2015review}%
  \BibitemOpen
  \bibfield  {author} {\bibinfo {author} {\bibfnamefont {S.~O.}\ \bibnamefont
  {Majekodunmi}},\ }\bibfield  {title} {\enquote {\bibinfo {title} {A review on
  centrifugation in the pharmaceutical industry},}\ }\href@noop {} {\bibfield
  {journal} {\bibinfo  {journal} {Am. J. Biomed. Eng}\ }\textbf {\bibinfo
  {volume} {5}},\ \bibinfo {pages} {67--78} (\bibinfo {year}
  {2015})}\BibitemShut {NoStop}%
\bibitem [{\citenamefont {Boisvert}\ \emph {et~al.}(2023)\citenamefont
  {Boisvert}, \citenamefont {Sadeghi}, \citenamefont {Rochefort},\ and\
  \citenamefont {Bazin}}]{boisvert2023axial}%
  \BibitemOpen
  \bibfield  {author} {\bibinfo {author} {\bibfnamefont {L.}~\bibnamefont
  {Boisvert}}, \bibinfo {author} {\bibfnamefont {M.}~\bibnamefont {Sadeghi}},
  \bibinfo {author} {\bibfnamefont {C.}~\bibnamefont {Rochefort}},\ and\
  \bibinfo {author} {\bibfnamefont {C.}~\bibnamefont {Bazin}},\ }\bibfield
  {title} {\enquote {\bibinfo {title} {Axial or turn-by-turn particle recovery
  in a spiral concentrator},}\ }\href@noop {} {\bibfield  {journal} {\bibinfo
  {journal} {CIM Journal}\ ,\ \bibinfo {pages} {1--14}} (\bibinfo {year}
  {2023})}\BibitemShut {NoStop}%
\bibitem [{\citenamefont {Berger}, \citenamefont {Talbot},\ and\ \citenamefont
  {Yao}(1983)}]{berger1983flow}%
  \BibitemOpen
  \bibfield  {author} {\bibinfo {author} {\bibfnamefont {S.}~\bibnamefont
  {Berger}}, \bibinfo {author} {\bibfnamefont {a.~L.}\ \bibnamefont {Talbot}},\
  and\ \bibinfo {author} {\bibfnamefont {L.}~\bibnamefont {Yao}},\ }\bibfield
  {title} {\enquote {\bibinfo {title} {Flow in curved pipes},}\ }\href@noop {}
  {\bibfield  {journal} {\bibinfo  {journal} {Annual review of fluid
  mechanics}\ }\textbf {\bibinfo {volume} {15}},\ \bibinfo {pages} {461--512}
  (\bibinfo {year} {1983})}\BibitemShut {NoStop}%
\bibitem [{\citenamefont {Ito}(1987)}]{ito1987flow}%
  \BibitemOpen
  \bibfield  {author} {\bibinfo {author} {\bibfnamefont {H.}~\bibnamefont
  {Ito}},\ }\bibfield  {title} {\enquote {\bibinfo {title} {Flow in curved
  pipes},}\ }\href@noop {} {\bibfield  {journal} {\bibinfo  {journal} {JSME
  international journal}\ }\textbf {\bibinfo {volume} {30}},\ \bibinfo {pages}
  {543--552} (\bibinfo {year} {1987})}\BibitemShut {NoStop}%
\bibitem [{\citenamefont {Germano}(1982)}]{germano1982effect}%
  \BibitemOpen
  \bibfield  {author} {\bibinfo {author} {\bibfnamefont {M.}~\bibnamefont
  {Germano}},\ }\bibfield  {title} {\enquote {\bibinfo {title} {On the effect
  of torsion on a helical pipe flow},}\ }\href@noop {} {\bibfield  {journal}
  {\bibinfo  {journal} {Journal of Fluid Mechanics}\ }\textbf {\bibinfo
  {volume} {125}},\ \bibinfo {pages} {1--8} (\bibinfo {year}
  {1982})}\BibitemShut {NoStop}%
\bibitem [{\citenamefont {Wang}(1981)}]{wang1981low}%
  \BibitemOpen
  \bibfield  {author} {\bibinfo {author} {\bibfnamefont {C.}~\bibnamefont
  {Wang}},\ }\bibfield  {title} {\enquote {\bibinfo {title} {On the
  low-reynolds-number flow in a helical pipe},}\ }\href@noop {} {\bibfield
  {journal} {\bibinfo  {journal} {Journal of Fluid Mechanics}\ }\textbf
  {\bibinfo {volume} {108}},\ \bibinfo {pages} {185--194} (\bibinfo {year}
  {1981})}\BibitemShut {NoStop}%
\bibitem [{\citenamefont {Meng}\ \emph {et~al.}(2023)\citenamefont {Meng},
  \citenamefont {Gao}, \citenamefont {Wei}, \citenamefont {Zhao}, \citenamefont
  {Cui}, \citenamefont {Shen},\ and\ \citenamefont
  {Song}}]{meng2023particulate}%
  \BibitemOpen
  \bibfield  {author} {\bibinfo {author} {\bibfnamefont {L.}~\bibnamefont
  {Meng}}, \bibinfo {author} {\bibfnamefont {S.}~\bibnamefont {Gao}}, \bibinfo
  {author} {\bibfnamefont {D.}~\bibnamefont {Wei}}, \bibinfo {author}
  {\bibfnamefont {Q.}~\bibnamefont {Zhao}}, \bibinfo {author} {\bibfnamefont
  {B.}~\bibnamefont {Cui}}, \bibinfo {author} {\bibfnamefont {Y.}~\bibnamefont
  {Shen}},\ and\ \bibinfo {author} {\bibfnamefont {Z.}~\bibnamefont {Song}},\
  }\bibfield  {title} {\enquote {\bibinfo {title} {Particulate flow modelling
  in a spiral separator by using the eulerian multi-fluid vof approach},}\
  }\href@noop {} {\bibfield  {journal} {\bibinfo  {journal} {International
  Journal of Mining Science and Technology}\ }\textbf {\bibinfo {volume}
  {33}},\ \bibinfo {pages} {251--263} (\bibinfo {year} {2023})}\BibitemShut
  {NoStop}%
\bibitem [{\citenamefont {Sudikondala}\ \emph {et~al.}(2022)\citenamefont
  {Sudikondala}, \citenamefont {Mangadoddy}, \citenamefont {Kumar},
  \citenamefont {Tripathy},\ and\ \citenamefont
  {Yanamandra}}]{sudikondala2022cfd}%
  \BibitemOpen
  \bibfield  {author} {\bibinfo {author} {\bibfnamefont {P.}~\bibnamefont
  {Sudikondala}}, \bibinfo {author} {\bibfnamefont {N.}~\bibnamefont
  {Mangadoddy}}, \bibinfo {author} {\bibfnamefont {M.}~\bibnamefont {Kumar}},
  \bibinfo {author} {\bibfnamefont {S.~K.}\ \bibnamefont {Tripathy}},\ and\
  \bibinfo {author} {\bibfnamefont {R.~M.}\ \bibnamefont {Yanamandra}},\
  }\bibfield  {title} {\enquote {\bibinfo {title} {Cfd modelling of spiral
  concentrator-prediction of comprehensive fluid flow field and particle
  segregation},}\ }\href@noop {} {\bibfield  {journal} {\bibinfo  {journal}
  {Minerals Engineering}\ }\textbf {\bibinfo {volume} {183}},\ \bibinfo {pages}
  {107570} (\bibinfo {year} {2022})}\BibitemShut {NoStop}%
\bibitem [{\citenamefont {Stokes}\ \emph {et~al.}(2013)\citenamefont {Stokes},
  \citenamefont {Duffy}, \citenamefont {Wilson},\ and\ \citenamefont
  {Tronnolone}}]{stokes2013thin}%
  \BibitemOpen
  \bibfield  {author} {\bibinfo {author} {\bibfnamefont {Y.~M.}\ \bibnamefont
  {Stokes}}, \bibinfo {author} {\bibfnamefont {B.~R.}\ \bibnamefont {Duffy}},
  \bibinfo {author} {\bibfnamefont {S.~K.}\ \bibnamefont {Wilson}},\ and\
  \bibinfo {author} {\bibfnamefont {H.}~\bibnamefont {Tronnolone}},\ }\bibfield
   {title} {\enquote {\bibinfo {title} {Thin-film flow in helically wound
  rectangular channels with small torsion},}\ }\href@noop {} {\bibfield
  {journal} {\bibinfo  {journal} {Physics of Fluids}\ }\textbf {\bibinfo
  {volume} {25}} (\bibinfo {year} {2013})}\BibitemShut {NoStop}%
\bibitem [{\citenamefont {Arnold}, \citenamefont {Stokes},\ and\ \citenamefont
  {Green}(2015)}]{arnold2015thin}%
  \BibitemOpen
  \bibfield  {author} {\bibinfo {author} {\bibfnamefont {D.}~\bibnamefont
  {Arnold}}, \bibinfo {author} {\bibfnamefont {Y.}~\bibnamefont {Stokes}},\
  and\ \bibinfo {author} {\bibfnamefont {J.}~\bibnamefont {Green}},\ }\bibfield
   {title} {\enquote {\bibinfo {title} {Thin-film flow in helically-wound
  rectangular channels of arbitrary torsion and curvature},}\ }\href@noop {}
  {\bibfield  {journal} {\bibinfo  {journal} {Journal of Fluid Mechanics}\
  }\textbf {\bibinfo {volume} {764}},\ \bibinfo {pages} {76--94} (\bibinfo
  {year} {2015})}\BibitemShut {NoStop}%
\bibitem [{\citenamefont {Arnold}, \citenamefont {Stokes},\ and\ \citenamefont
  {Green}(2017)}]{arnold2017thin}%
  \BibitemOpen
  \bibfield  {author} {\bibinfo {author} {\bibfnamefont {D.}~\bibnamefont
  {Arnold}}, \bibinfo {author} {\bibfnamefont {Y.}~\bibnamefont {Stokes}},\
  and\ \bibinfo {author} {\bibfnamefont {J.}~\bibnamefont {Green}},\ }\bibfield
   {title} {\enquote {\bibinfo {title} {Thin-film flow in helically wound
  shallow channels of arbitrary cross-sectional shape},}\ }\href@noop {}
  {\bibfield  {journal} {\bibinfo  {journal} {Physics of Fluids}\ }\textbf
  {\bibinfo {volume} {29}} (\bibinfo {year} {2017})}\BibitemShut {NoStop}%
\bibitem [{\citenamefont {Zhou}\ \emph {et~al.}(2005)\citenamefont {Zhou},
  \citenamefont {Dupuy}, \citenamefont {Bertozzi},\ and\ \citenamefont
  {Hosoi}}]{zhou2005theory}%
  \BibitemOpen
  \bibfield  {author} {\bibinfo {author} {\bibfnamefont {J.}~\bibnamefont
  {Zhou}}, \bibinfo {author} {\bibfnamefont {B.}~\bibnamefont {Dupuy}},
  \bibinfo {author} {\bibfnamefont {A.}~\bibnamefont {Bertozzi}},\ and\
  \bibinfo {author} {\bibfnamefont {A.}~\bibnamefont {Hosoi}},\ }\bibfield
  {title} {\enquote {\bibinfo {title} {Theory for shock dynamics in
  particle-laden thin films},}\ }\href@noop {} {\bibfield  {journal} {\bibinfo
  {journal} {Physical review letters}\ }\textbf {\bibinfo {volume} {94}},\
  \bibinfo {pages} {117803} (\bibinfo {year} {2005})}\BibitemShut {NoStop}%
\bibitem [{\citenamefont {Cook}(2008)}]{cook2008theory}%
  \BibitemOpen
  \bibfield  {author} {\bibinfo {author} {\bibfnamefont {B.~P.}\ \bibnamefont
  {Cook}},\ }\bibfield  {title} {\enquote {\bibinfo {title} {Theory for
  particle settling and shear-induced migration in thin-film liquid flow},}\
  }\href@noop {} {\bibfield  {journal} {\bibinfo  {journal} {Physical Review
  E}\ }\textbf {\bibinfo {volume} {78}},\ \bibinfo {pages} {045303} (\bibinfo
  {year} {2008})}\BibitemShut {NoStop}%
\bibitem [{\citenamefont {Murisic}\ \emph {et~al.}(2013)\citenamefont
  {Murisic}, \citenamefont {Pausader}, \citenamefont {Peschka},\ and\
  \citenamefont {Bertozzi}}]{murisic2013dynamics}%
  \BibitemOpen
  \bibfield  {author} {\bibinfo {author} {\bibfnamefont {N.}~\bibnamefont
  {Murisic}}, \bibinfo {author} {\bibfnamefont {B.}~\bibnamefont {Pausader}},
  \bibinfo {author} {\bibfnamefont {D.}~\bibnamefont {Peschka}},\ and\ \bibinfo
  {author} {\bibfnamefont {A.~L.}\ \bibnamefont {Bertozzi}},\ }\bibfield
  {title} {\enquote {\bibinfo {title} {Dynamics of particle settling and
  resuspension in viscous liquid films},}\ }\href@noop {} {\bibfield  {journal}
  {\bibinfo  {journal} {Journal of Fluid Mechanics}\ }\textbf {\bibinfo
  {volume} {717}},\ \bibinfo {pages} {203--231} (\bibinfo {year}
  {2013})}\BibitemShut {NoStop}%
\bibitem [{\citenamefont {Wong}, \citenamefont {Lindstrom},\ and\ \citenamefont
  {Bertozzi}(2019)}]{wong2019fast}%
  \BibitemOpen
  \bibfield  {author} {\bibinfo {author} {\bibfnamefont {J.}~\bibnamefont
  {Wong}}, \bibinfo {author} {\bibfnamefont {M.}~\bibnamefont {Lindstrom}},\
  and\ \bibinfo {author} {\bibfnamefont {A.~L.}\ \bibnamefont {Bertozzi}},\
  }\bibfield  {title} {\enquote {\bibinfo {title} {Fast equilibration dynamics
  of viscous particle-laden flow in an inclined channel},}\ }\href@noop {}
  {\bibfield  {journal} {\bibinfo  {journal} {Journal of Fluid Mechanics}\
  }\textbf {\bibinfo {volume} {879}},\ \bibinfo {pages} {28--53} (\bibinfo
  {year} {2019})}\BibitemShut {NoStop}%
\bibitem [{\citenamefont {Mirzaeian}, \citenamefont {Testik},\ and\
  \citenamefont {Alba}(2020)}]{mirzaeian2020bidensity}%
  \BibitemOpen
  \bibfield  {author} {\bibinfo {author} {\bibfnamefont {N.}~\bibnamefont
  {Mirzaeian}}, \bibinfo {author} {\bibfnamefont {F.}~\bibnamefont {Testik}},\
  and\ \bibinfo {author} {\bibfnamefont {K.}~\bibnamefont {Alba}},\ }\bibfield
  {title} {\enquote {\bibinfo {title} {Bidensity particle-laden exchange flows
  in a vertical duct},}\ }\href@noop {} {\bibfield  {journal} {\bibinfo
  {journal} {Journal of Fluid Mechanics}\ }\textbf {\bibinfo {volume} {891}},\
  \bibinfo {pages} {A18} (\bibinfo {year} {2020})}\BibitemShut {NoStop}%
\bibitem [{\citenamefont {Revay}\ and\ \citenamefont
  {Higdon}(1992)}]{revay1992numerical}%
  \BibitemOpen
  \bibfield  {author} {\bibinfo {author} {\bibfnamefont {J.}~\bibnamefont
  {Revay}}\ and\ \bibinfo {author} {\bibfnamefont {J.}~\bibnamefont {Higdon}},\
  }\bibfield  {title} {\enquote {\bibinfo {title} {Numerical simulation of
  polydisperse sedimentation: equal-sized spheres},}\ }\href@noop {} {\bibfield
   {journal} {\bibinfo  {journal} {Journal of Fluid Mechanics}\ }\textbf
  {\bibinfo {volume} {243}},\ \bibinfo {pages} {15--32} (\bibinfo {year}
  {1992})}\BibitemShut {NoStop}%
\bibitem [{\citenamefont {Leighton}\ and\ \citenamefont
  {Acrivos}(1987)}]{leighton1987shear}%
  \BibitemOpen
  \bibfield  {author} {\bibinfo {author} {\bibfnamefont {D.}~\bibnamefont
  {Leighton}}\ and\ \bibinfo {author} {\bibfnamefont {A.}~\bibnamefont
  {Acrivos}},\ }\bibfield  {title} {\enquote {\bibinfo {title} {The
  shear-induced migration of particles in concentrated suspensions},}\
  }\href@noop {} {\bibfield  {journal} {\bibinfo  {journal} {Journal of Fluid
  Mechanics}\ }\textbf {\bibinfo {volume} {181}},\ \bibinfo {pages} {415--439}
  (\bibinfo {year} {1987})}\BibitemShut {NoStop}%
\bibitem [{\citenamefont {Kanehl}\ and\ \citenamefont
  {Stark}(2015)}]{kanehl2015hydrodynamic}%
  \BibitemOpen
  \bibfield  {author} {\bibinfo {author} {\bibfnamefont {P.}~\bibnamefont
  {Kanehl}}\ and\ \bibinfo {author} {\bibfnamefont {H.}~\bibnamefont {Stark}},\
  }\bibfield  {title} {\enquote {\bibinfo {title} {Hydrodynamic segregation in
  a bidisperse colloidal suspension in microchannel flow: A theoretical
  study},}\ }\href@noop {} {\bibfield  {journal} {\bibinfo  {journal} {The
  Journal of chemical physics}\ }\textbf {\bibinfo {volume} {142}} (\bibinfo
  {year} {2015})}\BibitemShut {NoStop}%
\bibitem [{\citenamefont {Nott}\ and\ \citenamefont
  {Brady}(1994)}]{nott1994pressure}%
  \BibitemOpen
  \bibfield  {author} {\bibinfo {author} {\bibfnamefont {P.~R.}\ \bibnamefont
  {Nott}}\ and\ \bibinfo {author} {\bibfnamefont {J.~F.}\ \bibnamefont
  {Brady}},\ }\bibfield  {title} {\enquote {\bibinfo {title} {Pressure-driven
  flow of suspensions: simulation and theory},}\ }\href@noop {} {\bibfield
  {journal} {\bibinfo  {journal} {Journal of Fluid Mechanics}\ }\textbf
  {\bibinfo {volume} {275}},\ \bibinfo {pages} {157--199} (\bibinfo {year}
  {1994})}\BibitemShut {NoStop}%
\bibitem [{\citenamefont {Howard}, \citenamefont {Maxey},\ and\ \citenamefont
  {Gallier}(2022)}]{howard2022bidisperse}%
  \BibitemOpen
  \bibfield  {author} {\bibinfo {author} {\bibfnamefont {A.~A.}\ \bibnamefont
  {Howard}}, \bibinfo {author} {\bibfnamefont {M.~R.}\ \bibnamefont {Maxey}},\
  and\ \bibinfo {author} {\bibfnamefont {S.}~\bibnamefont {Gallier}},\
  }\bibfield  {title} {\enquote {\bibinfo {title} {Bidisperse supension balance
  model},}\ }\href@noop {} {\bibfield  {journal} {\bibinfo  {journal} {Physical
  Review Fluids}\ }\textbf {\bibinfo {volume} {7}},\ \bibinfo {pages} {124301}
  (\bibinfo {year} {2022})}\BibitemShut {NoStop}%
\bibitem [{\citenamefont {Lee}, \citenamefont {Wong},\ and\ \citenamefont
  {Bertozzi}(2015)}]{lee2015equilibrium}%
  \BibitemOpen
  \bibfield  {author} {\bibinfo {author} {\bibfnamefont {S.}~\bibnamefont
  {Lee}}, \bibinfo {author} {\bibfnamefont {J.}~\bibnamefont {Wong}},\ and\
  \bibinfo {author} {\bibfnamefont {A.~L.}\ \bibnamefont {Bertozzi}},\
  }\href@noop {} {\emph {\bibinfo {title} {Equilibrium theory of bidensity
  particle-laden flows on an incline}}}\ (\bibinfo  {publisher} {Springer},\
  \bibinfo {year} {2015})\BibitemShut {NoStop}%
\bibitem [{\citenamefont {Wang}\ and\ \citenamefont
  {Bertozzi}(2014)}]{wang2014shock}%
  \BibitemOpen
  \bibfield  {author} {\bibinfo {author} {\bibfnamefont {L.}~\bibnamefont
  {Wang}}\ and\ \bibinfo {author} {\bibfnamefont {A.~L.}\ \bibnamefont
  {Bertozzi}},\ }\bibfield  {title} {\enquote {\bibinfo {title} {Shock
  solutions for high concentration particle-laden thin films},}\ }\href@noop {}
  {\bibfield  {journal} {\bibinfo  {journal} {SIAM Journal on Applied
  Mathematics}\ }\textbf {\bibinfo {volume} {74}},\ \bibinfo {pages} {322--344}
  (\bibinfo {year} {2014})}\BibitemShut {NoStop}%
\bibitem [{\citenamefont {Krieger}\ and\ \citenamefont
  {Dougherty}(1959)}]{krieger1959mechanism}%
  \BibitemOpen
  \bibfield  {author} {\bibinfo {author} {\bibfnamefont {I.~M.}\ \bibnamefont
  {Krieger}}\ and\ \bibinfo {author} {\bibfnamefont {T.~J.}\ \bibnamefont
  {Dougherty}},\ }\bibfield  {title} {\enquote {\bibinfo {title} {A mechanism
  for non-newtonian flow in suspensions of rigid spheres},}\ }\href@noop {}
  {\bibfield  {journal} {\bibinfo  {journal} {Transactions of the Society of
  Rheology}\ }\textbf {\bibinfo {volume} {3}},\ \bibinfo {pages} {137--152}
  (\bibinfo {year} {1959})}\BibitemShut {NoStop}%
\bibitem [{\citenamefont {Morris}\ and\ \citenamefont
  {Boulay}(1999)}]{morris1999curvilinear}%
  \BibitemOpen
  \bibfield  {author} {\bibinfo {author} {\bibfnamefont {J.~F.}\ \bibnamefont
  {Morris}}\ and\ \bibinfo {author} {\bibfnamefont {F.}~\bibnamefont
  {Boulay}},\ }\bibfield  {title} {\enquote {\bibinfo {title} {Curvilinear
  flows of noncolloidal suspensions: The role of normal stresses},}\
  }\href@noop {} {\bibfield  {journal} {\bibinfo  {journal} {Journal of
  rheology}\ }\textbf {\bibinfo {volume} {43}},\ \bibinfo {pages} {1213--1237}
  (\bibinfo {year} {1999})}\BibitemShut {NoStop}%
\bibitem [{\citenamefont {Ovarlez}, \citenamefont {Bertrand},\ and\
  \citenamefont {Rodts}(2006)}]{ovarlez2006local}%
  \BibitemOpen
  \bibfield  {author} {\bibinfo {author} {\bibfnamefont {G.}~\bibnamefont
  {Ovarlez}}, \bibinfo {author} {\bibfnamefont {F.}~\bibnamefont {Bertrand}},\
  and\ \bibinfo {author} {\bibfnamefont {S.}~\bibnamefont {Rodts}},\ }\bibfield
   {title} {\enquote {\bibinfo {title} {Local determination of the constitutive
  law of a dense suspension of noncolloidal particles through magnetic
  resonance imaging},}\ }\href@noop {} {\bibfield  {journal} {\bibinfo
  {journal} {Journal of rheology}\ }\textbf {\bibinfo {volume} {50}},\ \bibinfo
  {pages} {259--292} (\bibinfo {year} {2006})}\BibitemShut {NoStop}%
\bibitem [{\citenamefont {Tripathi}\ and\ \citenamefont
  {Acrivos}(1999)}]{tripathi1999viscous}%
  \BibitemOpen
  \bibfield  {author} {\bibinfo {author} {\bibfnamefont {A.}~\bibnamefont
  {Tripathi}}\ and\ \bibinfo {author} {\bibfnamefont {A.}~\bibnamefont
  {Acrivos}},\ }\bibfield  {title} {\enquote {\bibinfo {title} {Viscous
  resuspension in a bidensity suspension},}\ }\href@noop {} {\bibfield
  {journal} {\bibinfo  {journal} {International journal of multiphase flow}\
  }\textbf {\bibinfo {volume} {25}},\ \bibinfo {pages} {1--14} (\bibinfo {year}
  {1999})}\BibitemShut {NoStop}%
\bibitem [{\citenamefont {Schaflinger}, \citenamefont {Acrivos},\ and\
  \citenamefont {Zhang}(1990)}]{schaflinger1990viscous}%
  \BibitemOpen
  \bibfield  {author} {\bibinfo {author} {\bibfnamefont {U.}~\bibnamefont
  {Schaflinger}}, \bibinfo {author} {\bibfnamefont {A.}~\bibnamefont
  {Acrivos}},\ and\ \bibinfo {author} {\bibfnamefont {K.}~\bibnamefont
  {Zhang}},\ }\bibfield  {title} {\enquote {\bibinfo {title} {Viscous
  resuspension of a sediment within a laminar and stratified flow},}\
  }\href@noop {} {\bibfield  {journal} {\bibinfo  {journal} {International
  Journal of Multiphase Flow}\ }\textbf {\bibinfo {volume} {16}},\ \bibinfo
  {pages} {567--578} (\bibinfo {year} {1990})}\BibitemShut {NoStop}%
\bibitem [{\citenamefont {Phillips}\ \emph {et~al.}(1992)\citenamefont
  {Phillips}, \citenamefont {Armstrong}, \citenamefont {Brown}, \citenamefont
  {Graham},\ and\ \citenamefont {Abbott}}]{phillips1992constitutive}%
  \BibitemOpen
  \bibfield  {author} {\bibinfo {author} {\bibfnamefont {R.~J.}\ \bibnamefont
  {Phillips}}, \bibinfo {author} {\bibfnamefont {R.~C.}\ \bibnamefont
  {Armstrong}}, \bibinfo {author} {\bibfnamefont {R.~A.}\ \bibnamefont
  {Brown}}, \bibinfo {author} {\bibfnamefont {A.~L.}\ \bibnamefont {Graham}},\
  and\ \bibinfo {author} {\bibfnamefont {J.~R.}\ \bibnamefont {Abbott}},\
  }\bibfield  {title} {\enquote {\bibinfo {title} {A constitutive equation for
  concentrated suspensions that accounts for shear-induced particle
  migration},}\ }\href@noop {} {\bibfield  {journal} {\bibinfo  {journal}
  {Physics of Fluids A: Fluid Dynamics}\ }\textbf {\bibinfo {volume} {4}},\
  \bibinfo {pages} {30--40} (\bibinfo {year} {1992})}\BibitemShut {NoStop}%
\bibitem [{\citenamefont {Sierou}\ and\ \citenamefont
  {Brady}(2004)}]{sierou2004shear}%
  \BibitemOpen
  \bibfield  {author} {\bibinfo {author} {\bibfnamefont {A.}~\bibnamefont
  {Sierou}}\ and\ \bibinfo {author} {\bibfnamefont {J.~F.}\ \bibnamefont
  {Brady}},\ }\bibfield  {title} {\enquote {\bibinfo {title} {Shear-induced
  self-diffusion in non-colloidal suspensions},}\ }\href@noop {} {\bibfield
  {journal} {\bibinfo  {journal} {Journal of fluid mechanics}\ }\textbf
  {\bibinfo {volume} {506}},\ \bibinfo {pages} {285--314} (\bibinfo {year}
  {2004})}\BibitemShut {NoStop}%
\bibitem [{\citenamefont {Murisic}\ \emph {et~al.}(2011)\citenamefont
  {Murisic}, \citenamefont {Ho}, \citenamefont {Hu}, \citenamefont {Latterman},
  \citenamefont {Koch}, \citenamefont {Lin}, \citenamefont {Mata},\ and\
  \citenamefont {Bertozzi}}]{murisic2011particle}%
  \BibitemOpen
  \bibfield  {author} {\bibinfo {author} {\bibfnamefont {N.}~\bibnamefont
  {Murisic}}, \bibinfo {author} {\bibfnamefont {J.}~\bibnamefont {Ho}},
  \bibinfo {author} {\bibfnamefont {V.}~\bibnamefont {Hu}}, \bibinfo {author}
  {\bibfnamefont {P.}~\bibnamefont {Latterman}}, \bibinfo {author}
  {\bibfnamefont {T.}~\bibnamefont {Koch}}, \bibinfo {author} {\bibfnamefont
  {K.}~\bibnamefont {Lin}}, \bibinfo {author} {\bibfnamefont {M.}~\bibnamefont
  {Mata}},\ and\ \bibinfo {author} {\bibfnamefont {A.}~\bibnamefont
  {Bertozzi}},\ }\bibfield  {title} {\enquote {\bibinfo {title} {Particle-laden
  viscous thin-film flows on an incline: Experiments compared with a theory
  based on shear-induced migration and particle settling},}\ }\href@noop {}
  {\bibfield  {journal} {\bibinfo  {journal} {Physica D: Nonlinear Phenomena}\
  }\textbf {\bibinfo {volume} {240}},\ \bibinfo {pages} {1661--1673} (\bibinfo
  {year} {2011})}\BibitemShut {NoStop}%
\bibitem [{\citenamefont {Manoussaki}\ and\ \citenamefont
  {Chadwick}(2000)}]{manoussaki2000effects}%
  \BibitemOpen
  \bibfield  {author} {\bibinfo {author} {\bibfnamefont {D.}~\bibnamefont
  {Manoussaki}}\ and\ \bibinfo {author} {\bibfnamefont {R.~S.}\ \bibnamefont
  {Chadwick}},\ }\bibfield  {title} {\enquote {\bibinfo {title} {Effects of
  geometry on fluid loading in a coiled cochlea},}\ }\href@noop {} {\bibfield
  {journal} {\bibinfo  {journal} {SIAM Journal on Applied Mathematics}\
  }\textbf {\bibinfo {volume} {61}},\ \bibinfo {pages} {369--386} (\bibinfo
  {year} {2000})}\BibitemShut {NoStop}%
\bibitem [{\citenamefont {Hill}\ and\ \citenamefont
  {Stokes}(2018)}]{hill2018note}%
  \BibitemOpen
  \bibfield  {author} {\bibinfo {author} {\bibfnamefont {J.~M.}\ \bibnamefont
  {Hill}}\ and\ \bibinfo {author} {\bibfnamefont {Y.~M.}\ \bibnamefont
  {Stokes}},\ }\bibfield  {title} {\enquote {\bibinfo {title} {A note on
  navier--stokes equations with nonorthogonal coordinates},}\ }\href@noop {}
  {\bibfield  {journal} {\bibinfo  {journal} {The ANZIAM Journal}\ }\textbf
  {\bibinfo {volume} {59}},\ \bibinfo {pages} {335--348} (\bibinfo {year}
  {2018})}\BibitemShut {NoStop}%
\bibitem [{\citenamefont {Stokes}\ \emph {et~al.}(2004)\citenamefont {Stokes},
  \citenamefont {Wilson}, \citenamefont {Duffy}, \citenamefont {Behnia},
  \citenamefont {Lin},\ and\ \citenamefont {McBain}}]{stokes2004thin}%
  \BibitemOpen
  \bibfield  {author} {\bibinfo {author} {\bibfnamefont {Y.}~\bibnamefont
  {Stokes}}, \bibinfo {author} {\bibfnamefont {S.}~\bibnamefont {Wilson}},
  \bibinfo {author} {\bibfnamefont {B.}~\bibnamefont {Duffy}}, \bibinfo
  {author} {\bibfnamefont {M.}~\bibnamefont {Behnia}}, \bibinfo {author}
  {\bibfnamefont {W.}~\bibnamefont {Lin}},\ and\ \bibinfo {author}
  {\bibfnamefont {G.}~\bibnamefont {McBain}},\ }\bibfield  {title} {\enquote
  {\bibinfo {title} {Thin-film flow in open helically-wound channels},}\ }in\
  \href@noop {} {\emph {\bibinfo {booktitle} {15th Australasian Fluid Mechanics
  Conference}}}\ (\bibinfo {organization} {Citeseer},\ \bibinfo {year} {2004})\
  pp.\ \bibinfo {pages} {13--17}\BibitemShut {NoStop}%
\bibitem [{jxs()}]{jxscmachine2024table}%
  \BibitemOpen
  \href@noop {} {\enquote {\bibinfo {title} {{JXSC} mine machinery factory
  spiral chute technical parameters table},}\ }\bibinfo {howpublished}
  {\url{https://www.jxscmachine.com/gravity-separator/spiral-chute/}},\
  \bibinfo {note} {accessed on: 2024-01-08}\BibitemShut {NoStop}%
\end{thebibliography}

\end{document}